\documentclass[
  10pt,
  aps,
  prd,
  twocolumn,
  longbibliography,
  superscriptaddress,
  nofootinbib,
  amsmath,
  amssymb,
  eqsecnum,
  floatfix
]{revtex4-2}

\usepackage[utf8]{inputenc}
\usepackage{mathtools}
\usepackage{appendix}
\usepackage{amsthm}
\usepackage{graphicx}
\usepackage{xcolor}
\usepackage{physics}
\usepackage{braket}
\usepackage{dcolumn}
\usepackage{bm}
\usepackage{hyperref}

\hypersetup{
  colorlinks,
  linkcolor=blue,
  citecolor=blue,
  urlcolor=blue
}

\let\originalleft\left
\let\originalright\right
\renewcommand{\left}{\mathopen{}\mathclose\bgroup\originalleft}
\renewcommand{\right}{\aftergroup\egroup\originalright}

\theoremstyle{plain}

\theoremstyle{remark}

\DeclareMathOperator{\sinc}{sinc}

\makeatletter

\newcommand{\ringplus}{\mathbin{\text{\@ringplus}}}

\newcommand{\@ringplus}{%
  \ooalign{\hidewidth\raise1.3ex\hbox{\tiny$\circ$}\hidewidth\cr$\m@th+$\cr}%
}

\newcommand{\ringminus}{\mathbin{\text{\@ringminus}}}

\newcommand{\@ringminus}{%
  \ooalign{\hidewidth\raise0.9ex\hbox{\tiny$\circ$}\hidewidth\cr$\m@th-$\cr}%
}
\makeatother

\newcommand{\bvec}[1]{\bm{#1}}

\DeclareFontFamily{U}{wncy}{}
\DeclareFontShape{U}{wncy}{m}{n}{<->wncyr10}{}
\DeclareSymbolFont{mcy}{U}{wncy}{m}{n}
\DeclareMathSymbol{\Sh}{\mathord}{mcy}{"58}

\newcommand{\reals}[0]{\mathbb{R}}

\newcommand{\integers}[0]{\mathbb{Z}}

\providecommand{\op}[1]{}
\renewcommand{\op}[1]{\hat{#1}}

\allowdisplaybreaks[1]

\begin{document}
\title{Organised multi-scale correlations in interacting QFT with Shannon wavelets}

\newcommand{\affmq}{School of Mathematical and Physical Sciences, Macquarie University, Sydney, NSW 2109, Australia}

\newcommand{\affrmit}{%
Centre for Quantum Computer Performance and Integration, School of Science, RMIT University, Melbourne, Victoria 3000, Australia}

\newcommand{\affuwaterloophys}{Department of Physics and Astronomy, University of Waterloo, Waterloo, ON, N2L 1W6, Canada}

\newcommand{\affuwaterloomath}{Department of Applied Mathematics, University of Waterloo, Waterloo, Ontario, N2L 3G1, Canada}

\newcommand{\affUwaterlooIQC}{Institute for Quantum Computing, University of Waterloo, Waterloo, Ontario, N2L 3G1, Canada}

\newcommand{\affperimeter}{Perimeter Institute for Theoretical Physics, Waterloo, Ontario, N2L 2Y5, Canada}

\author{Dominic G. Lewis}
 \email{dominic.lewis@student.rmit.edu.au}
\affiliation{\affrmit}%
\author{Nicholas Funai}
\affiliation{\affrmit}%

\author{\v{S}imon Vedl}
\affiliation{\affmq}%

\author{Daniel George}
\affiliation{\affmq}%

\author{Gavin K. Brennen}
 \affiliation{\affmq}%

\author{Fil Simovic}
\affiliation{\affmq}
\affiliation{\affuwaterloophys}
 
\author{Achim Kempf}
\affiliation{\affuwaterloomath}
\affiliation{\affUwaterlooIQC}
\affiliation{\affperimeter}

\author{Nicolas C. Menicucci}
 \email{nicolas.menicucci@rmit.edu.au}
 \affiliation{\affrmit}%

\date{30 November 2026}
\begin{abstract}
    Discrete representations make quantum fields more tractable, but a conventional lattice can obscure continuum symmetries. Shannon sampling supplies an exact continuous--discrete equivalence for fields with a global ultraviolet cut-off, which excludes all momenta above a fixed scale. A generic field is not bandlimited, and imposing such a cut-off removes part of the theory one wishes to describe. What is missing is an exact discrete representation that retains all momentum scales and makes clear how those scales communicate once interactions are present. We construct such a representation using Shannon wavelets, whose non-overlapping Fourier supports partition the field into bandlimited components. Each component has its own sampling lattice, so a field in $d$ spatial dimensions is represented exactly by discrete position and scale labels without imposing either an ultraviolet or an infrared cut-off. For quadratic Hamiltonians, derivative operators, time evolution, and two-point correlations separate exactly by wavelet scale. A quartic self-interaction couples these otherwise independent sectors, but momentum conservation imposes selection rules on the allowed scale combinations. Surprisingly, the exact interacting two-point function remains scale diagonal; cross-scale structure first appears in higher-order correlators. Shannon wavelets therefore expose a hierarchy of multiscale communication that is hidden in both the continuum field and its conventional lattice representations.
\end{abstract}
\maketitle

\section{Introduction}
Quantum field theory is formulated using fields that vary continuously over spacetime, yet many of its analytical and numerical methods replace that continuum with a lattice or restrict the available momenta \cite{UpdatedPeskin2018,UpdatedNg2009}. These methods are useful, but they introduce a tension. A fixed lattice selects a preferred set of spatial points and can obscure the continuous symmetries of the original theory, while a hard ultraviolet cut-off removes high-frequency degrees of freedom \cite{UpdatedMuroya2003, UpdatedDrell1974, UpdatedDrell21976}. This tension is especially important in settings where a minimum resolvable length is expected but a smooth spacetime description must still emerge \cite{UpdatedHossenfelder2013,kempf1999generalized,UpdatedPye2015,Kempf2000,ChatwinDaviesNaturalCovariant,UpdatedGrimmer2022,UpdatedGaray1995,UpdatedMead1964}. The problem is therefore not simply how to discretise a quantum field but how to do so without needlessly giving up its continuous description or its symmetries.

Sampling theory provides an exact bridge between continuous and discrete descriptions, but only for a restricted class of fields. A globally bandlimited field can be reconstructed perfectly from sufficiently dense samples, and its lattice representation remains isomorphic to the continuous field without taking a continuum limit \cite{UpdatedShannon1948,Kempf2000,UpdatedPye2015,Pye2023,UpdatedLewis2023}. A generic field, however, has unbounded momentum support. Imposing a global bandlimit in order to use this correspondence changes the field one wishes to describe. What remains unclear is whether an non-band-limited field can be represented exactly by discrete degrees of freedom while retaining all of its momentum scales, and, crucially, how those scales couple when the field self-interacts. This gap matters because interactions between resolutions are central to renormalisation, multiscale simulation, and proposals in which scale behaves as an additional geometric direction. An exact scale-resolved representation would make it possible to distinguish structures imposed by a regulator from those already present in the continuum theory.

Wavelets offer a natural way to construct such a representation. Rather than discard momenta above a single cut-off, a wavelet transform resolves a field into components associated with different positions and frequency scales \cite{Daubechies1992,cattani2008shannon}. This resembles the organisation of degrees of freedom in the renormalisation group, where the behaviour of a theory is studied across characteristic momentum or length scales \cite{Wilson1974,Cardy1996,Goldenfeld1992,BergesTetradisWetterich2002,Polchinski1984,Wetterich1993}. This relationship has been explored in the past \cite{altaisky2016,battle1999}, however the important difference is that a complete wavelet transform retains every scale at once. It can therefore expose the multiscale structure of a field without yet integrating out, truncating, or approximating any part of it.

Wavelet methods have already found several uses in quantum field theory. Daubechies wavelets provide discrete multiresolution representations with natural volume and resolution truncations \cite{BulutPolyzou2013,MichlinPolyzouBulut2017}, while continuous wavelet transforms have been applied to interacting Green functions and scale-dependent formulations of field theories \cite{Altaisky2013}. Wavelet bases have also been proposed for multiscale quantum simulation \cite{Brennen2015}, used to study correlations and entanglement across resolutions \cite{GeorgeBrennen2022}, and related directly to entanglement-renormalisation constructions \cite{EvenblyWhite2016}. Infinite-range wavelets can reproduce continuum spectra over a finite energy range in lattice renormalisation transformations \cite{Fries2019}.
More recently, a momentum-space Daubechies basis has been applied nonperturbatively to $(1+1)$-dimensional $\phi^4$ theory, providing systematic infrared and ultraviolet truncations and increasingly accurate access to the strong-coupling phase transition as the resolution is increased \cite{Basak2026}. Together, these results show that scale is not merely an auxiliary wavelet label; it is a useful way to organise the physical and computational degrees of freedom of a field.

In terms of quantum field theory with bandpass cut-offs in the spirit of Shannon wavelets, recent work by Thiemann develops a Dirichlet-Shannon multiresolution framework and illustrates Dirichlet-kernel renormalisation for a free scalar field on a spatial circle \cite{Thiemann2023}. Its follow up applies this Dirichlet-kernel Hamiltonian renormalization scheme to a self-interacting scalar bosonic QFT in 1+1 spacetime dimensions at finite volume, showing that it recovers the theory's known continuum fixed point \cite{ZarateThiemann2025}. 

The wavelet scale label has also motivated connections between wavelets and holography. Exact holographic mappings and related constructions transform a lower-dimensional system into a multiscale representation with an additional direction, where correlations and entanglement may admit a geometric interpretation \cite{LeeQi2016,SinghBrennen2016,ChingHuaLee2017}. In recent work, we used a continuous wavelet transform to construct a scale-dependent bulk description of quadratic quantum field theories and defined an information metric from local multiscale correlations using the Petz--R\'enyi mutual information \cite{VedlBrennen2026}. For the free scalar and fermionic theories considered there, the resulting geometries are warped Euclidean anti-de Sitter spaces and reduce to Euclidean AdS in the massless ground-state case \cite{Maldacena1998,AmmonErdmenger2015}. The present work begins from the same observation that a wavelet transform supplies an additional scale coordinate but asks a different question. We study the exact discrete structure and interaction pattern of the transformed field itself, without assuming that its position--scale representation has a gravitational interpretation.

Shannon wavelets are particularly useful for this purpose because their Fourier supports are bounded and do not overlap between scales. At scale $n$, a Shannon wavelet contains only spatial momenta satisfying
\begin{equation}
2^n\Omega<|k|<2^{n+1}\Omega,
\end{equation}
and neighbouring scales partition momentum space into distinct dyadic bands. The parameter $\Omega$ fixes the widths of these bands but does not act as a cut-off. Allowing $n\in\mathbb Z$ covers the momentum line apart from the single point $k=0$: increasingly fine bands accumulate near zero momentum, while increasingly broad bands extend without bound into the ultraviolet. The complete field therefore needs neither an ultraviolet nor an infrared cut-off, even though each wavelet component is bandlimited between two scale-dependent frequencies. Shannon sampling can then be applied independently to every component. The result is an exact discrete representation in which each scale has its own natural sampling lattice. Discreteness enters here through a complete choice of basis rather than through a regulator; cut-offs arise only if the range of wavelet scales is subsequently truncated. In this way, the continuous--discrete correspondence for globally bandlimited fields \cite{Kempf2000,UpdatedPye2015,UpdatedLewis2023} extends scale by scale to fields with unbounded momentum support.

This construction adds a discrete scale label to the usual position labels. A field in $d$ spatial dimensions may therefore be represented by a lattice carrying $d$ position coordinates and one scale coordinate, while time remains unchanged. In one spatial dimension, the scale and translation indices $(n,j)$ can also be combined into a single coordinate on the dyadic rationals $\eta\in\mathbb D^\times$, a dense subset of the reals \cite{Gouvea1997,Koblitz1984,Grafakos2014}. Throughout this work we use both of these representations to connect $(n,j)$ representations of the discrete wavelets used commonly in wavelet-based quantum field theory work \cite{VedlBrennen2026,GeorgeBrennen2022} to the representations used commonly throughout the broader established Shannon wavelets literature \cite{Chui1992,Daubechies1992,Mallat2008,Meyer1993,Wojtaszczyk1997,Mallat1989}. 

We calculate the discrete equivalents to the derivatives on each wavelet scale as matrices and show that these matrices have no elements between distinct wavelet scales. This is a calculation that has close ties to the SLAC or bandlimited derivatives \cite{UpdatedDrell1974,UpdatedDrell21976,UpdatedLewis2023} and is known in existing wavelet literature \cite{cattani2008shannon} yet its usefulness is rather understated. Here, we apply these derivative matrices in place of the continuous wave operator to give an exact discrete Hamiltonian for the scalar field that separates entirely into different scales. We then show how this scale separation manifests in the two-point functions of the free field.

Self-interactions change this structure. We study a $(1+1)$-dimensional scalar field with a quartic interaction and show that products of four Shannon wavelets couple scale sectors that are independent at quadratic order. These couplings are described by a quartic wavelet-overlap tensor, but they are not arbitrary and instead obey distinct selection rules depending on the difference in wavelet scales.

The coupling in the Hamiltonian leads to a further question: which correlations can actually reveal communication between scales? We answer this question by determining the scale dependencies on the two-point vacuum correlation function as well as its higher counterparts involving higher order powers of the fields. Surprisingly, the exact interacting two-point function remains zero between distinct Shannon scales and one must go to higher order correlations in order to get non-zero interscale communication.

The remainder of this paper is organised as follows. In Sec.~\ref{sec:waveletdecomposition} we introduce the Shannon wavelet decomposition and develop a sampling theory for square-integrable functions without bounded total Fourier support. We derive discrete forms of the $L^2$ inner product and spatial derivatives. In Sec.~\ref{sec:waveletscalarfield} we apply these tools to the free scalar field and determine its commutation relations, Hamiltonian, time evolution, and scale-resolved two-point correlations. In Sec.~\ref{sec:dyadicrepfreescalarfield} we combine the position and scale indices into the dyadic-rational coordinate $\eta$. In Sec.~\ref{sec:interactingwaveletfield} we introduce the quartic interaction and derive the momentum-support selection rules for cross-scale couplings. In Sec.~\ref{sec:interactingcorrelationfunctionswavelets} we prove that the interacting two-point function remains scale diagonal  
 and extend the selection rule to general $N$-point functions. We conclude in Sec.~\ref{sec:waveletsoutlook} with possible applications to renormalisation, simulation, and emergent geometry.

\section{Shannon wavelet decomposition \label{sec:waveletdecomposition}}
\begin{figure}
    \centering
    \includegraphics[width=\linewidth]{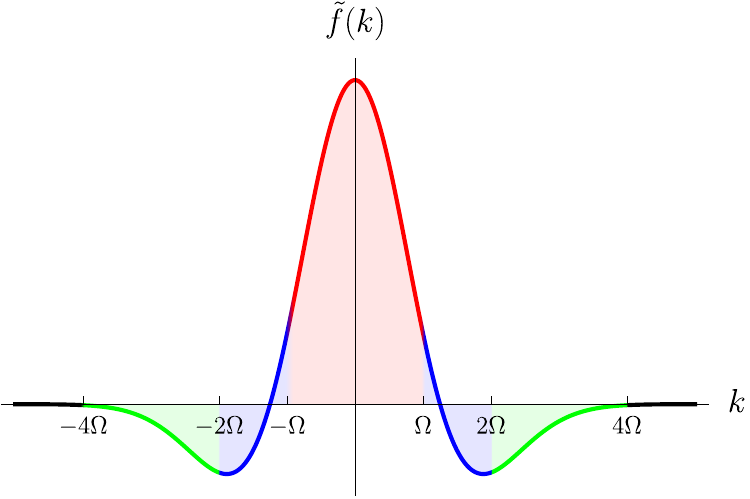}
    \caption{The decomposition of some test signal into its low frequency (red), medium frequency (blue) and high frequency (green) regions.}
    \label{fig:Shannonwaveletdecomposition1}
\end{figure}

Before considering the full decomposition of a function with Shannon wavelets, it is useful to first visualise this process for a simple signal and few wavelet scales. As seen in Fig.~\ref{fig:Shannonwaveletdecomposition1}, we have separated a signal into three distinct regions (labeled by colour), and each region, with the exception of the center in this case, is itself characterised by an ultraviolet cut-off at $2^{n+1}\Omega$ and an infrared cut-off at $2^n \Omega$.
The center region has only an ultraviolet cut-off at $\Omega$ and is called bandlimited---see \cite{UpdatedLewis2023, UpdatedPye2015, kempf1999generalized} for sampling theories and treatment of bandlimited quantum field theories. For the purpose of this work, we will largely omit the center bandlimited component of the frequency decomposition. We will see later that this has little effect on the wavelet decompositions of both functions and quantum fields with the caveat that the zero-frequency mode, or constant, function becomes excluded.

\subsection{Decomposition and sampling theory with discrete Shannon wavelets}
We use the Fourier convention
\begin{align}
    \tilde f(k)&\coloneqq\int_\reals dx\,f(x)e^{-ikx},&
    f(x)&=\int_\reals\frac{dk}{2\pi}\tilde f(k)e^{ikx}.
\end{align}
For $\Omega>0$ and $n,j\in\integers$, define
\begin{align}
    a_n&\coloneqq 2^n\Omega,&
    h_n&\coloneqq\frac{\pi}{a_n}=2^{-n}\Delta x,&
    x_{nj}&\coloneqq h_n(j+1/2),
    \label{waveletscaledictionary}
\end{align}
where $\Delta x\coloneqq\pi/\Omega$. Here $a_n$ is the lower band edge, $h_n$ is the sampling spacing, and $x_{nj}$ includes the half-site shift at every scale. The label $n$ denotes scale and $j$ denotes translation. On the wavelets and their coefficients we write $n$ as an upper index and $j$ as a lower index; in $a_n$, $h_n$ and $x_{nj}$ these labels appear as subscripts.

\begin{figure}[t]
    \centering
    \includegraphics[width=\linewidth]{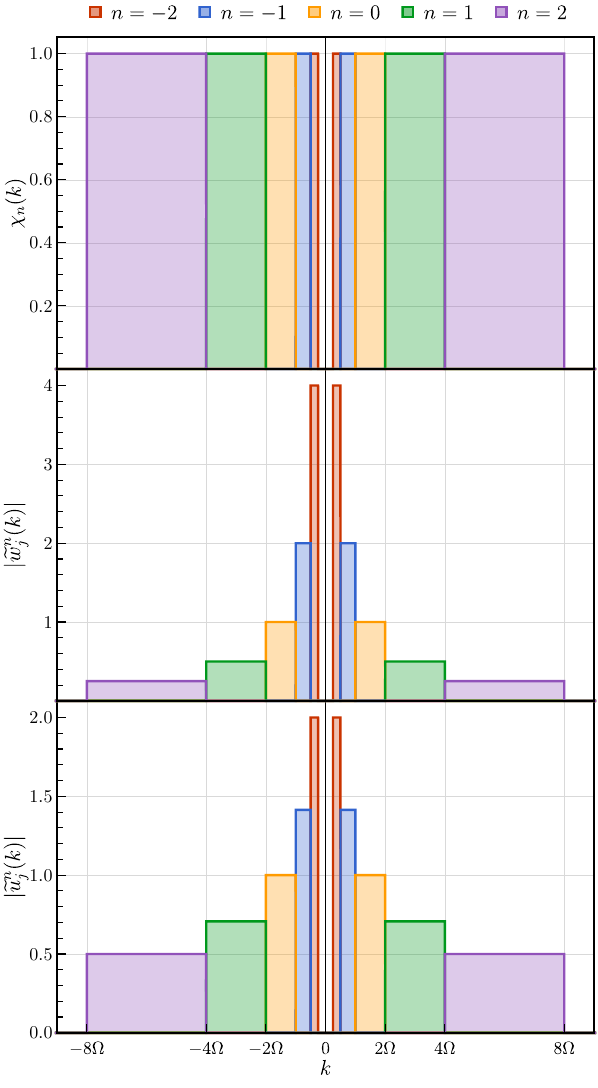}
    \caption{Representative Shannon bands and their normalisation, with $\Omega=\pi$. The top panel shows the band indicators $\chi_n(k)$; each band has width $a_n=2^n\Omega$. The middle and bottom panels show $|\widetilde w^n_j(k)|=h_n\chi_n(k)$ and $|\widetilde u^n_j(k)|=\sqrt{h_n}\chi_n(k)$, respectively, using the normalisations defined below. The plotted scales are $n=-2,\ldots,2$. We can see clearly that as $n$ runs through the negative integers, the band support approaches the origin but its decreasing width never allows it to cross it. Extending the bands over all $n\in\integers$ thus gives support on $\reals\setminus\{0\}$, up to band edges.}
    \label{fig:waveletstophats}
\end{figure}

With the top hat convention
\begin{align}
    \Pi(z)\coloneqq
    \begin{cases}
        1,&|z|<1/2,\\
        0,&\text{otherwise},
    \end{cases}
\end{align}
we define the band indicator used throughout:
\begin{align}
    \chi_n(k)&\coloneqq\mathbf{1}_{\{a_n<|k|<2a_n\}}
    \nonumber\\
    &=\Pi\left(\frac{k}{4a_n}\right)-\Pi\left(\frac{k}{2a_n}\right).
    \label{waveletbandindicator}
\end{align}
The top hat representation holds almost everywhere (a.e.); values at band edges are set by the strict inequalities in the first line and do not affect the Fourier integrals. Thus,
\begin{align}
    \chi_n(k)\chi_m(k)&=\delta_{nm}\chi_n(k),&
    \sum_{n\in\integers}\chi_n(k)&=1\quad\text{a.e.}
    \label{bandprojectoridentities}
\end{align}
The frequency decomposition in Fig.~\ref{fig:Shannonwaveletdecomposition1} is
\begin{align}
    \tilde f(k)&=\tilde f(k)\Pi\left(\frac{k}{2\Omega}\right)
        +\sum_{n=0}^\infty\tilde f(k)\chi_n(k).
    \label{bandedfunctiondecomposition}
\end{align}
Extending $n$ over all integers replaces the central low-frequency band by an infinite sequence of wavelet bands. Grouping each band component explicitly, we have
\begin{align}
    \tilde f(k)=\sum_{n\in\integers}\underbrace{\tilde f(k)\chi_n(k)}_{\widetilde{f^n}(k)}\quad\text{a.e.}
\end{align}
Here $\widetilde{f^n}(k)$ is the Fourier transform of the band projection $f^n$ defined in Eq.~\eqref{fndef}.
The parameter $\Omega$ sets the band widths; it is not a cut-off when all scales are retained. Fig.~\ref{fig:waveletstophats} shows representative bands and the corresponding Fourier magnitudes for both wavelet normalisations.

\begin{figure}[!hb]
    \centering
    \includegraphics[width=\linewidth]{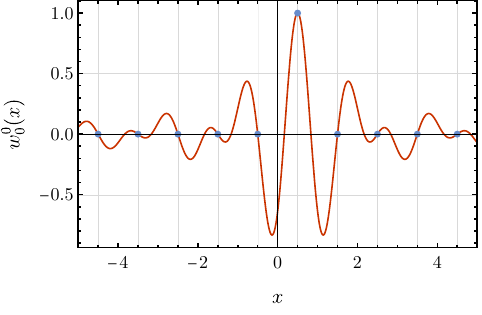}
    \caption{Plot of the profile of the cardinal Shannon wavelet $w^0_0(x)$ for scale $n=0$ and translation $j=0$. Here, $\Omega=\pi$ so that $h_0=1$ and the sampling sites are the integers translated by $1/2$. The wavelet is one at $x_{00}=1/2$ and zero at all other sampling sites. The cardinal normalisation is retained here to show the direct sampling interpretation.}
    \label{fig:shannonwaveletplot}
\end{figure}

To retain the direct connection with Shannon sampling, define the \emph{cardinal} wavelet
\begin{align}
    w^n_j(x)&\coloneqq 2\sinc\bigl(2a_n(x-x_{nj})\bigr)
                  -\sinc\bigl(a_n(x-x_{nj})\bigr).
    \label{cardinalshannonwavelet}
\end{align}
Here $\sinc z\coloneqq\sin z/z$, with its continuous value at zero. These are the sampling kernels used in this section:
\begin{align}
    w^n_j(x_{n\ell})=\delta_{j\ell}.
    \label{cardinalwaveletsamples}
\end{align}
The half-site shifts make the sampling sites at distinct scales distinct. An example of the cardinal wavelet and its sampling values is shown in Fig.~\ref{fig:shannonwaveletplot}.

The Fourier transform and overlap of the cardinal wavelets are
\begin{align}
    \widetilde w^n_j(k)&=h_ne^{-ikx_{nj}}\chi_n(k),
    \label{cardinalwaveletfourier}\\
    \int_\reals dx\,\overline{w^n_j(x)}w^m_\ell(x)
       &=h_n\delta_{nm}\delta_{j\ell}.
    \label{waveletoverlap}
\end{align}
Thus the cardinal wavelets are orthogonal, with squared norm $h_n$. We also define the \emph{orthonormal} wavelets here:
\begin{align}
    u^n_j(x)&\coloneqq\frac{w^n_j(x)}{\sqrt{h_n}},
    \nonumber\\
    \widetilde u^n_j(k)&=\sqrt{h_n}e^{-ikx_{nj}}\chi_n(k),
    \label{orthonormalshannonwavelet}\\
    \int_\reals dx\,\overline{u^n_j(x)}u^m_\ell(x)
       &=\delta_{nm}\delta_{j\ell}.
    \label{orthonormalwaveletoverlap}
\end{align}
For a square-integrable function $f$, the coefficient dictionary is
\begin{subequations}
\begin{align}
    f^n_j&\coloneqq\frac{1}{h_n}\int_\reals dx\,\overline{w^n_j(x)}f(x)
       =f^n(x_{nj}),\\
    F^n_j&\coloneqq\int_\reals dx\,\overline{u^n_j(x)}f(x)
       =\sqrt{h_n}f^n_j,
    \label{waveletcoefficientdictionary}
\end{align}
\end{subequations}
where $f^n$ is the band projection in Eq.~\eqref{fndef}. Both expansions reconstruct the same function:
\begin{align}
    f(x)=\sum_{n,j\in\integers}f^n_jw^n_j(x)
        =\sum_{n,j\in\integers}F^n_ju^n_j(x).
    \label{twonormalizationreconstruction}
\end{align}
Lower-case $f^n_j$ denotes a literal sample of the projected function; $F^n_j$ denotes an orthonormal coefficient. We retain the cardinal representation below to exhibit the sampling interpretation. From Sec.~\ref{sec:waveletscalarfield} onward, all discrete quantum-field variables and overlap tensors use $u$, with no further change of normalization.

\subsubsection{Wavelet decomposition of functions}

A function that has an ultraviolet cut-off at $\Omega$ can be sampled perfectly onto a lattice using Shannon's sampling formula as
\begin{align}
    f(x) = \sum_{j\in\integers}f(x_j)S^\Omega_j(x),
\end{align}
where 
\begin{align}
    S_j^\Omega(x) \coloneqq \sinc\left(\pi\frac{x-x_j}{\Delta x}\right)
\end{align}
is the sinc function given by the support of $f$ in the Fourier domain, with the base sampling sites $x_j\coloneqq j\Delta x$.
Through Shannon's sampling theorem, we can draw an equivalence between continuous and discrete representations of bandlimited functions~\cite{UpdatedShannon1948}. Using Shannon wavelets, we extend this notion to functions without a hard ultraviolet cut-off~\cite{cattani2008shannon}.

Consider a function whose Fourier transform is supported over the full real line. We can decompose such a function into a sum of bandlimited component functions as shown in Eq.~\eqref{bandedfunctiondecomposition}, using bands shown in Fig.~\ref{fig:waveletstophats}. In other words, we can separate a non-band-limited function $f$ into an infinite sum of bandlimited ones at different scales. At each scale, such continuous function will have a reconstruction formula analogous to Shannon reconstruction
\begin{align}
    f^n(x) = \sum_{j\in\integers}f^n_jw^n_j(x),
\end{align}
where $f^n(x)$ is given by
\begin{align}
    f^n(x) 
    :=
    \int_\mathbb{R}\frac{dk}{2\pi}\tilde{f}(k)e^{ikx} \chi_n(k),
    \label{fndef}
\end{align}
and $f^n_j$ is given by
\begin{align}
    f^n_j
    &:= \frac{1}{h_n}
    \int_{\mathbb{R}}dxf(x)\overline{w^n_j(x)}
    \nonumber
\\
    &=
    f^n(x_{nj}).
\end{align}
Now, from the linearity of the Fourier transform, we can see that combining Eq.~\eqref{fndef} with Eq.~\eqref{bandedfunctiondecomposition} gives the following, where $\widetilde{f^\Omega}(k)\coloneqq\tilde f(k)\Pi(k/(2\Omega))$ denotes the central band:
\begin{align}
    f(x) 
    =&
    \int_\mathbb{R}\frac{dk}{2\pi}\tilde{f}(k)e^{ikx}
    \nonumber
\\*
    =&
    \int_\mathbb{R}\frac{dk}{2\pi}e^{ikx}\Bigl[\widetilde{f^\Omega}(k)+\sum_{n=0}^\infty\widetilde{f^n}(k)\Bigr]
    \nonumber
\\*
    =&
    f^\Omega(x) + \sum_{n=0}^\infty f^{n}(x),
\end{align}
where $f^{\Omega}$ is the component of $f$ that is $\Omega$-bandlimited.
\begin{align}
    f^\Omega (x) \coloneqq \int_{-\Omega}^\Omega\frac{dk}{2\pi}\tilde{f}(k)e^{ikx}.
\end{align}
Combining these allows us to extend the Shannon's sampling formula to functions that are not necessarily cut-off in frequency support. We have
\begin{align}
    f(x) = \sum_{j\in\integers}f^\Omega_jS^\Omega_j(x) + \sum_{n=0}^\infty \sum_{k\in\integers}f^n_kw^n_k(x).
    \label{extendedshannonsampling}
\end{align}
As noted previously, $\Omega$ is no longer a UV cut-off, instead it sets the width of the scale of the wavelets. Additionally, if one wanted to introduce a UV cut-off to a function decomposed in this way, they would simply need to truncate the sum over $n$, effectively setting a maximum scale to the wavelets.

The $L^2$ inner product becomes a weighted sum of products of cardinal samples. Orthogonality between the central sinc band and the bands with $n\geq0$, together with Eq.~\eqref{waveletoverlap}, gives
\begin{align}
    \int_\reals dx\,\overline{f(x)}g(x)
       &=\Delta x\sum_j\overline{f^\Omega_j}g^\Omega_j
       +\sum_{n=0}^\infty h_n\sum_j\overline{f^n_j}g^n_j.
    \label{waveletinnerproduct}
\end{align}
The factors $h_n$ are the sampling spacings and squared norms of the cardinal wavelets.

Now, instead of including the scale (sinc) function contribution to the scale decomposition of functions, it is also possible to simply allow the scale number $n$ to run into the negative integers as well as the positive ones. This approach replaces the center band in the frequency support of a function with an infinite number of pairs of increasingly thinning wavelet bands. This alternative deconstruction of a function's Fourier transform has the form
\begin{align}
    f(x) &= \int_\reals \frac{dk}{2\pi}\tilde{f}(k)e^{ikx}
    \nonumber
\\
    &= 
    \int_\reals \frac{dk}{2\pi}e^{ikx}\sum_{n\in\integers}\tilde{f}(k) \chi_n(k)
    \nonumber
\\
    &=
    \sum_{n\in\integers}f^n(x).
\end{align}
Consequently, we can rewrite the generalised sampling theorem as just an infinite sum of scale reconstructions.
\begin{align}
    f(x) = \sum_{n\in\integers}\sum_{k\in\integers}f^n_kw^n_k(x).
    \label{waveletsamplingalt}
\end{align}
It is important to note here that the forms of the wavelet functions themselves have not been changed by this alteration to reconstruction of $f$, we have simply included more of them.%
\footnote{It is also important to note that by replacing the center bandlimited component of the scale decomposition with more wavelets, we are omitting the center $k=0$ mode from the Fourier support of these functions. Such an omission acts to exclude the constant function from the support of the wavelet decomposition. However, for the purposes of this work when dealing with quantum fields and $L^2$ integrable functions, it turns out that this omission is of little import due to constant functions being omitted from many quantum field theoretic calculations. That being said, it is possible to reintroduce the constant function by appending the zero mode frequency to the decomposition of momentum space.}\ %
Consequently, properties of the wavelet functions such as their orthogonality  are unaffected by this choice. As such, we can immediately write the $L^2$ inner product of two functions as
\begin{align}
    \int_\reals dx\,\overline{f(x)}g(x)
       &=\sum_{n\in\integers}h_n\sum_j\overline{f^n_j}g^n_j
       \nonumber\\
       &=\sum_{n,j\in\integers}\overline{F^n_j}G^n_j.
    \label{shannonwaveletinnerprod}
\end{align}
It is also useful to note that this approach is conceptually equivalent to letting the scaling parameter $\Omega$ become infinitesimally small and then re-indexing the sum over $n$. However, this also highlights the important caveat that truncating the sum over $n$ in the negative direction now serves to act an infrared cut-off on the function space, analogous to how truncating $n$ in the positive direction imposes an ultraviolet cut-off to the function space.

\subsubsection{Derivatives}
To differentiate a function in the discrete representation, we need a matrix that acts on its wavelet coefficients. The cardinal and orthonormal representations of the wavelets give the same derivative operator. We first define its matrix elements by the overlap with a differentiated wavelet:
\begin{align}
    D^{(\ell)nm}_{jh}
       &\coloneqq\int_\reals dx\,\overline{u^n_j(x)}\partial_x^\ell u^m_h(x)
       \nonumber\\
       &=\frac{1}{\sqrt{h_nh_m}}\int_\reals dx\,
                   \overline{w^n_j(x)}\partial_x^\ell w^m_h(x).
    \label{waveletderivativematrix}
\end{align}
Here the row labels $(n,j)$ specify the output coefficient and the column labels $(m,h)$ specify the input coefficient. The derivative order is parenthesised: $D^{(2)}$ denotes a second derivative. Since differentiation multiplies a Fourier transform by $(ik)^\ell$, it preserves each band. Using Eq.~\eqref{bandprojectoridentities},
\begin{align}
    D^{(\ell)nm}_{jh}
       &=\sqrt{h_nh_m}\int_\reals\frac{dk}{2\pi}(ik)^\ell e^{ik(x_{nj}-x_{mh})}
       \nonumber\\*
       &\hspace{42mm}\times\chi_n(k)\chi_m(k)
       \nonumber\\*
       &=\delta_{nm}D^{(\ell)n}_{jh},\\*
    D^{(\ell)n}_{jh}
       &\coloneqq h_n\int_\reals\frac{dk}{2\pi}(ik)^\ell
          e^{ikh_n(j-h)}\chi_n(k).
    \label{gammacalc1}
\end{align}
Equivalently, the same-scale matrix elements are the dimensionless band integrals
\begin{align}
    D^{(\ell)n}_{jh}
       &=\frac{h_n^{-\ell}}{2\pi}
          \left(\int_{-2\pi}^{-\pi}+\int_\pi^{2\pi}\right)
          ds\,(is)^\ell e^{is(j-h)}.
    \label{dimensionlessderivativeintegral}
\end{align}
For completeness, integrating these band integrals by parts gives the general matrix element. With $d=j-h$ and nonnegative integer $\ell$,
\begin{align}
    &D^{(\ell)n}_{jh} = a_n^{\ell}\partial^{\ell}_x \left[ 2\sinc(2x) - \sinc(x) \right]_{x=\pi (j-h)}
    \nonumber\\
    &\quad=\frac{i^\ell}{2\pi h_n^\ell}
    \begin{cases}
       \displaystyle
       \frac{(2\pi)^{\ell+1}-\pi^{\ell+1}}{\ell+1}
          \bigl[1+(-1)^\ell\bigr],&d=0,\\[2mm]
       \begin{aligned}[b]
       &\displaystyle\sum_{r=0}^{\ell}
       \frac{(-1)^r\ell!}{(\ell-r)!(id)^{r+1}}
       \bigl[1-(-1)^{\ell-r}\bigr]\\
       &\qquad\times\bigl[(2\pi)^{\ell-r}-(-1)^d\pi^{\ell-r}\bigr]
       \end{aligned},&d\neq0.
    \end{cases}
    \label{gammaelldef}
\end{align}
For cardinal samples the matrix element is instead initially obtained with $1/h_n$ multiplying the cardinal overlap. This equals the orthonormal matrix element when $n=m$, and both vanish when $n\neq m$. Hence the same $D^{(\ell)n}$ acts on either set of coefficients:
\begin{align}
    (\partial^\ell f)^n_j&=\sum_h D^{(\ell)n}_{jh}f^n_h,&
    (\partial^\ell F)^n_j&=\sum_h D^{(\ell)n}_{jh}F^n_h,
    \label{waveletderivative1}
\end{align}
where $(\partial^\ell F)^n_j$ denotes the orthonormal coefficient of $\partial_x^\ell f$, so it equals $\sqrt{h_n}(\partial^\ell f)^n_j$. Reconstruction gives
\begin{align}
    \partial_x^\ell f(x)
       &=\sum_{n,j,h}D^{(\ell)n}_{jh}f^n_h w^n_j(x)
       \nonumber\\*
       &=\sum_{n,j,h}D^{(\ell)n}_{jh}F^n_h u^n_j(x).
    \label{waveletderivativedef}
\end{align}
These derivative identities apply when the indicated derivative belongs to $L^2$.

\begin{figure}[t]
    \centering
    \includegraphics[width=\linewidth]{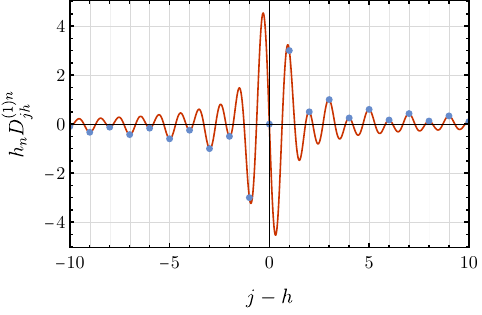}
    \caption{Plot of the dimensionless first-derivative matrix elements $h_nD^{(1)n}_{hj}$ as a function of $j-h$, evaluated on the integers. %
    }
    \label{fig:waveletderivatives1}
\end{figure}

\begin{figure}[t]
    \centering
    \includegraphics[width=\linewidth]{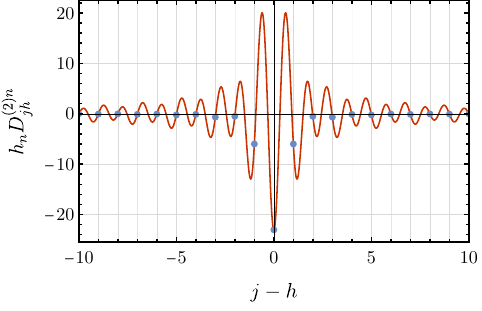}
    \caption{Plot of the dimensionless second-derivative matrix elements $h_n^2D^{(2)n}_{hj}$ as a function of $j-h$, evaluated on the integers. It is important to note that these entries are all negative, as given by the explicit Shannon-wavelet matrix elements. 
    }
    \label{fig:waveletderivatives2}
\end{figure}

For the first and second derivatives, the band integrals give
\begin{subequations}
\begin{align}
    D^{(1)n}_{jh}
       &=\frac{1}{h_n}
       \begin{cases}
           0,&j=h,\\
           \dfrac{2-(-1)^{j-h}}{j-h},&j\neq h,
       \end{cases}\\
    D^{(2)n}_{jh}
       &=\frac{1}{h_n^2}
       \begin{cases}
           -\dfrac{7\pi^2}{3},&j=h,\\
           -\dfrac{2(2-(-1)^{j-h})}{(j-h)^2},&j\neq h.
       \end{cases}
\end{align}
\end{subequations}
We can note immediately that $D^{(1)n}$ is antisymmetric and $D^{(2)n}$ is symmetric. The dimensionless elements $h_n^\ell D^{(\ell)n}_{hj}$ are plotted in Figs.~\ref{fig:waveletderivatives1} and~\ref{fig:waveletderivatives2}, respectively as functions of the difference in lattice index $j-h$.

It is interesting to note that the first derivative matrix elements are strictly positive for positive separation $j-h$ and negative for negative separation while the matrix elements for the second derivative are all strictly negative regardless of separation.
With the normalization and derivative conventions fixed, we now apply the orthonormal representation to quantum fields.

\section{Wavelet based sampling theory for the scalar field  \label{sec:waveletscalarfield}}

Starting with the free scalar field in one dimension \cite{UpdatedPeskin2018, UpdatedNg2009}, given by
\begin{align}
    \hat{\phi}(x)
    = 
    \int_\mathbb{R}\frac{dk}{2\pi}\frac{1}{\sqrt{2\omega_k}}\left(\hat{a}_k + \hat{a}_{-k}^\dagger\right)e^{ikx},
    \label{KGphi2}
\end{align}
where
\begin{align}
    \omega_k \coloneqq \sqrt{k^2+m^2},
\end{align}
we use the orthonormal wavelets of Eq.~\eqref{orthonormalshannonwavelet} for the quantum-field expansion:
\begin{align}
    \hat{\phi}(x)&=\sum_{n,j\in\integers}\hat Q^n_j u^n_j(x),&
    \hat{\pi}(x)&=\sum_{n,j\in\integers}\hat P^n_j u^n_j(x).
    \label{decompositionoffield}
\end{align}
The discrete field variables are the smeared operators
\begin{align}
    \hat Q^n_j&\coloneqq\int_\reals dx\,\overline{u^n_j(x)}\hat\phi(x)
       \nonumber\\
       &=\sqrt{h_n}\int_\reals\frac{dk}{2\pi}
           \hat{\tilde\phi}(k)e^{ikx_{nj}}\chi_n(k),\\
    \hat P^n_j&\coloneqq\int_\reals dx\,\overline{u^n_j(x)}\hat\pi(x)
       \nonumber\\
       &=\sqrt{h_n}\int_\reals\frac{dk}{2\pi}
           \hat{\tilde\pi}(k)e^{ikx_{nj}}\chi_n(k).
    \label{canonicalwaveletmodes}
\end{align}
We reserve $\hat\phi^n(x)$ and $\hat\pi^n(x)$ for the continuous band-projected fields:
\begin{align}
    \hat\phi^n(x)&\coloneqq\int_\reals\frac{dk}{2\pi}
                   \hat{\tilde\phi}(k)e^{ikx}\chi_n(k),\\
    \hat\pi^n(x)&\coloneqq\int_\reals\frac{dk}{2\pi}
                   \hat{\tilde\pi}(k)e^{ikx}\chi_n(k).
    \label{bandprojectedfields}
\end{align}
Thus the sampling dictionary reads
\begin{align}
    \hat Q^n_j&=\sqrt{h_n}\,\hat\phi^n(x_{nj}),&
    \hat P^n_j&=\sqrt{h_n}\,\hat\pi^n(x_{nj}).
    \label{quantumsamplingdictionary}
\end{align}
The half-site shift in $x_{nj}=h_n(j+1/2)$ is retained in every Fourier phase. The operators $Q$ and $P$ are orthonormal coefficients; dividing them by $\sqrt{h_n}$ recovers the literal cardinal samples. Each projected field has support $a_n<|k|<2a_n$.

\begin{figure}
    \centering
    \includegraphics[width=\linewidth]{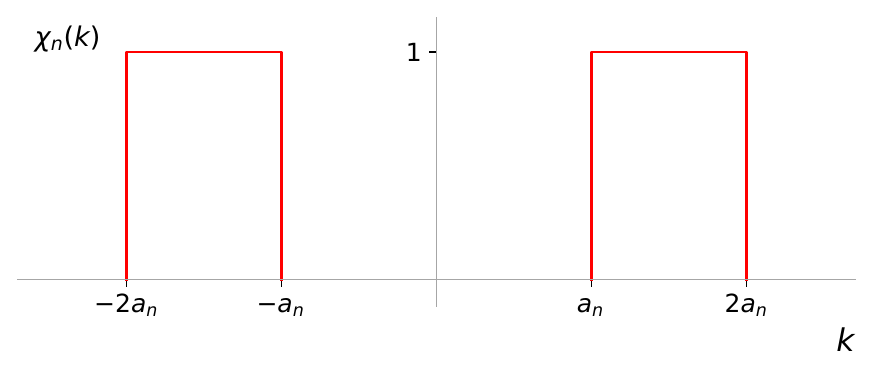}
    \caption{Plot of the Fourier support of $\hat{\phi}^n$ at a single wavelet scale $n$, shown by the band indicator $\chi_n(k)$. Here, the field can only produce modes of momentum within $a_n<|k|<2a_n$, as these bands specify the support for which $\op a$ and $\op a^\dagger$ live at this scale. The height is that of the indicator, not the Fourier amplitude of a wavelet.}
    \label{fig:fieldfourierdomain}
\end{figure}

In Fig.~\ref{fig:fieldfourierdomain} we can see the Fourier domain for $\hat{\phi}^n$.
Additionally, we can see that there is no overlap between any $\hat{\phi}^n$ for all $n\in \integers$ as expected through the definition of the Shannon wavelet decomposition given in the previous section.
Now, from Eq.~\eqref{decompositionoffield}, we can write the continuous scalar field as an infinite sum of the samples of its $\emph{bandlimited}$ (with UV and IR cut--offs) components from lattices of spacing $2^{-n}\frac{\pi}{\Omega}$. As there is no overlap of the field with itself at different scales, we can see that $\hat{\phi}^n$ and $\hat{\phi}^m$ are orthogonal and, as a result, commute with each other.

\subsection{Commutation relations}
The continuum equal-time relation $[\hat\phi(x),\hat\pi(y)]=i\delta(x-y)$ and orthonormality give
\begin{align}
    [\hat Q^n_j,\hat P^m_h]
       &=i\int_\reals dx\,\overline{u^n_j(x)}u^m_h(x)
        =i\delta_{nm}\delta_{jh},\\
    [\hat Q^n_j,\hat Q^m_h]&=[\hat P^n_j,\hat P^m_h]=0.
    \label{canonicalwaveletcommutators}
\end{align}
For the continuous projected fields, the same calculation uses $\chi_n(k)\chi_m(-k)=\delta_{nm}\chi_n(k)$:
\begin{align}
    [\hat\phi^n(x),\hat\pi^m(y)]
       &=i\delta_{nm}\int_\reals\frac{dk}{2\pi}e^{ik(x-y)}\chi_n(k)
       \nonumber\\*
       &=\frac{i\delta_{nm}}{h_n}\bigl[2\sinc(2a_nr)-\sinc(a_nr)\bigr],
    \label{waveletcommutator}
\end{align}
where $r=x-y$. Evaluating on the sampling sites yields
\begin{align}
    [\hat\phi^n(x_{nj}),\hat\pi^m(x_{mh})]
       =\frac{i}{h_n}\delta_{nm}\delta_{jh}.
\end{align}
The factors $\sqrt{h_n}$ in Eq.~\eqref{quantumsamplingdictionary} convert this sample commutator into the canonical one.

\begin{figure}
    \centering
    \includegraphics[width=\linewidth]{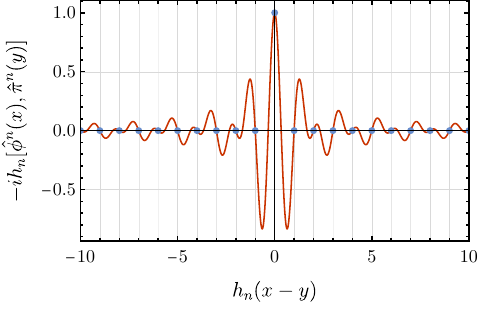}
    \caption{Plot of the commutation relation divided by $i$ between $\hat{\phi}^n(x)$ and $\hat{\pi}^n(y)$ as a function of $x-y$. Note that for this plot $\Omega=\pi$ and $n=2$, so that $h_2=1/4$. The band kernel $B_{a_2}(x-y)$ has height $1/h_2=4$ at coincidence and is zero at non-zero integer multiples of $h_2$.}
    \label{fig:waveletcommutator}
\end{figure}

A plot of the commutation relation given in Eq.~\eqref{waveletcommutator} can be seen in Fig.~\ref{fig:waveletcommutator}.
As such, the field operators at different scales and different points on the lattice commute with each other and fail to commute when they are at the same lattice point and at the same scale. On the appropriate sampling lattice, the band kernel in Eq.~\eqref{waveletcommutator} reduces to a Kronecker delta multiplied by $1/h_n$. However, shifting a single band into the ultraviolet does not recover a Dirac delta. To see this, write the kernel for the band $(a,2a)$ as
\begin{align}
    B_a(r)&\coloneqq\int_{a<|k|<2a}\frac{dk}{2\pi}e^{ikr}
       =\frac{\sin(2ar)-\sin(ar)}{\pi r}.
    \label{singlebandcommutatorkernel}
\end{align}
For any Schwartz test function $g$, its pairing with this kernel is
\begin{align}
    \int_\reals dr\,B_a(r)g(r)
       &=\int_{a<|k|<2a}\frac{dk}{2\pi}\tilde g(-k)
         \longrightarrow 0
         \quad(a\longrightarrow\infty),
\end{align}
since $\tilde g$ decays rapidly at large momentum. Thus $B_a$ tends to zero in the sense of tempered distributions, including when $\Omega\rightarrow\infty$ at fixed $n$. In order to recover the full Dirac delta, one must instead sum over all scales:
\begin{align}
    \sum_{n\in\integers}B_{a_n}(x-y)
       &=\int_\reals\frac{dk}{2\pi}e^{ik(x-y)}
         \sum_{n\in\integers}\chi_n(k)
       \nonumber\\
       &=\delta(x-y).
    \label{allscalecommutatorkernel}
\end{align}
This identity is understood distributionally and is consistent with the fact that the full Klein--Gordon field is a sum over all scales of the bandlimited fields $\hat{\phi}^n$ and $\hat{\pi}^n$.

Note that the other commutation relations between the fields follow the standard QFT conventions and as such all equal zero.

\subsection{Hamiltonian}
We consider the Hamiltonian for the free scalar field in one dimension
\begin{align}
    \hat{H}_{KG}
    \coloneqq
    \frac{1}{2}\int_{\mathbb{R}}dx\left[\left(\hat{\pi}(x)\right)^2-\hat{\phi}(x)\Delta \hat{\phi}(x) + m^2\left(\hat{\phi}(x)\right)^2\right].
    \label{KGHam2}
\end{align}
Using the orthonormal inner product in Eq.~\eqref{shannonwaveletinnerprod} and the derivative matrix in Eq.~\eqref{waveletderivativematrix}, the discrete Hamiltonian is
\begin{align}
    \hat H_{KG}
       &=\frac12\sum_{n,j}\left[(\hat P^n_j)^2+m^2(\hat Q^n_j)^2\right]
       \nonumber\\
       &\quad-\frac12\sum_{n,j,h}\hat Q^n_jD^{(2)n}_{jh}\hat Q^n_h.
    \label{waveletHamiltonian}
\end{align}
Accordingly, define
\begin{align}
    \hat H^n
       &\coloneqq\frac12\sum_j\left[(\hat P^n_j)^2+m^2(\hat Q^n_j)^2\right]
       \nonumber\\
       &\quad-\frac12\sum_{j,h}\hat Q^n_jD^{(2)n}_{jh}\hat Q^n_h,
    \label{discreteHamiltonians}\\
    \hat H_{KG}&=\sum_{n\in\integers}\hat H^n.
    \label{Hamiltoniandecomposition}
\end{align}
There is no overall $h_n$ factor for a scale Hamiltonian in canonical variables. Such a factor belongs to the cardinal-sample expression and is absorbed by $Q^n_j=\sqrt{h_n}\phi^n(x_{nj})$ and $P^n_j=\sqrt{h_n}\pi^n(x_{nj})$.

Now, as there is nothing that couples the terms in Eq.~\eqref{Hamiltoniandecomposition}, it is possible to consider this field as a collection of smaller fields that are completely isolated from each other.
Each $\hat{H}^n_{KG}$ describes a bandlimited (or bandpass) Klein--Gordon field with UV cut-off at $2^{n+1}\Omega$ and IR cut-off at $2^n\Omega$. Each of these fields have their own perceived minimum length scale and associated Nyquist lattices with spacings at %
$2^{-n}\frac{\pi}{\Omega}$, respectively.

If one wished to implement hard ultraviolet and/or infrared cut-offs to this theory, they would simply need to truncate the sum over $n$ in either direction at whatever scale they wish. Such a truncation from below sets the scale for the \emph{coarsest} lattice and a truncation from above sets the scale for the \emph{finest} lattice.

\begin{figure}
    \centering
    \includegraphics[width=\linewidth]{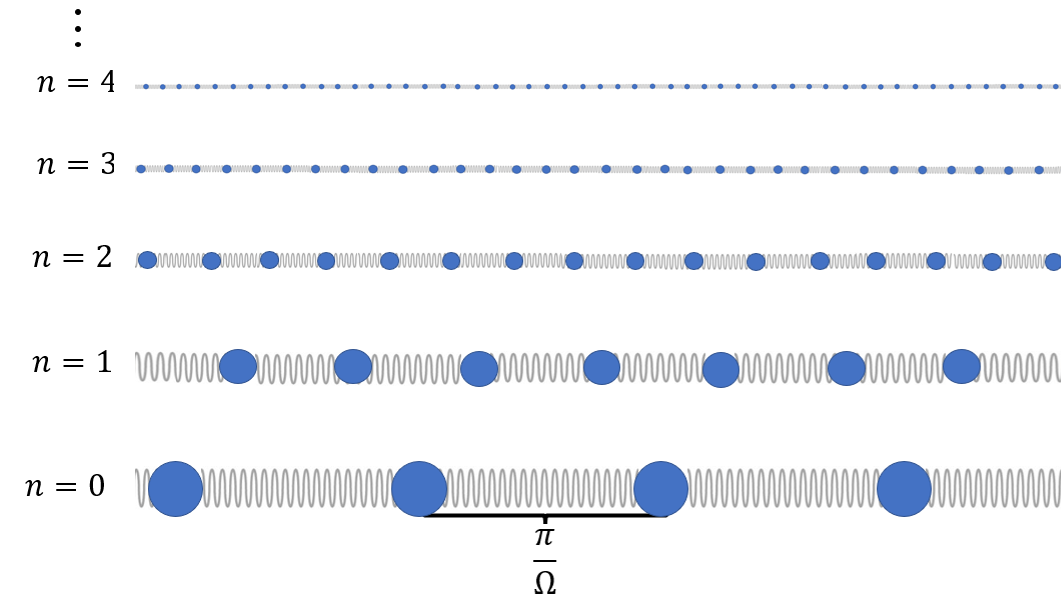}
    \caption{Graphical representation of a discrete 2+1 dimensional lattice field theory using the scale of the wavelet basis as the second spatial dimension. Note that the springs between the balls represent non-local couplings between harmonic oscillators at the lattice sites. These springs couple each oscillator in a layer to all others in the same line. Note that oscillators between layers are not coupled together. We can see that as $n$ increases so does the density of the lattice in each respective layer. The varying marker sizes are schematic: with the canonical variables used here, there is no overall scale-dependent prefactor multiplying each scale Hamiltonian.}
    \label{fig:waveletlattice}
\end{figure}

If one were to place these lattices on top of each other in a manner illustrated via Fig.~\ref{fig:waveletlattice}, they would construct what appears to be a  1+1\emph{+1} dimensional system where the additional dimension is the scale $n$. It is important to note, however, that this extra dimension along scale index is purely illustrative. In fact, given that there is no coupling between the scales under the Shannon wavelet decomposition, one cannot infer any geometry to such a higher dimensional lattice structure.
Instead, we simply use the picture of lattice layers stacked on top of one another as a means of interpreting co-ordinates and the separability of these scales. We will see in a later section that one can instead refer to the co-ordinate system of these lattices in terms of the dyadic rationals. We will also see that for theories of oder beyond quadratic such as those with interactions, the separability of the different Shannon scales is lost.

\subsection{Scale separation in time evolution and correlations}
\begin{figure}
    \centering
    \includegraphics[width=\linewidth]{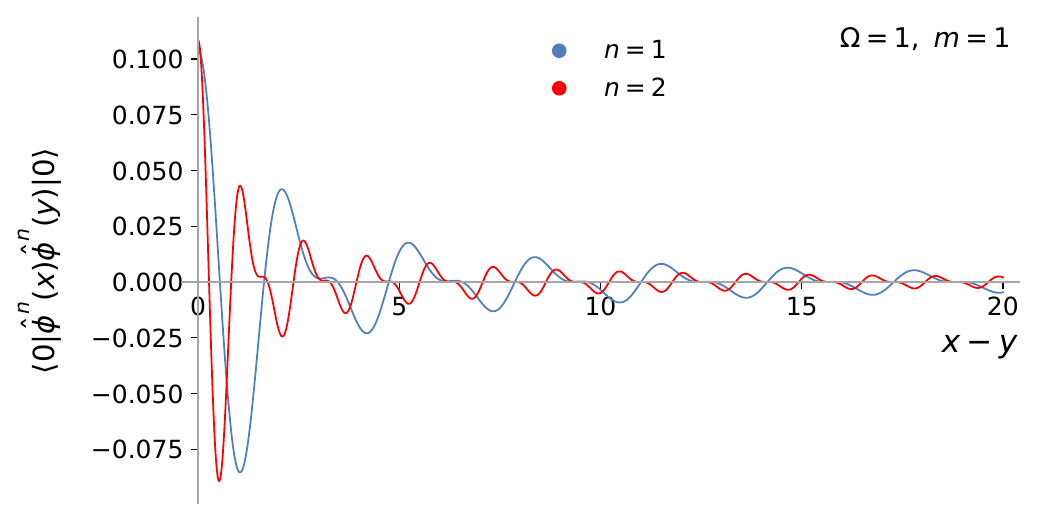}
    \caption{Plots of the equal-time correlation functions for the wavelet components $\hat{\phi}^n$ as functions of $x-y$ for $\Omega=m=1$. The blue plot is at $n=1$ and the red plot is at scale $n=2$. These are the band-field correlators of Eq.~\eqref{equaltimebandcorrelator}; multiplying by $h_n$ at the sampling sites gives the corresponding canonical $Q^n_j$ correlators.}
    \label{fig:waveletcorrelationfunction}
\end{figure}

\begin{figure}
    \centering
    \includegraphics[width=\linewidth]{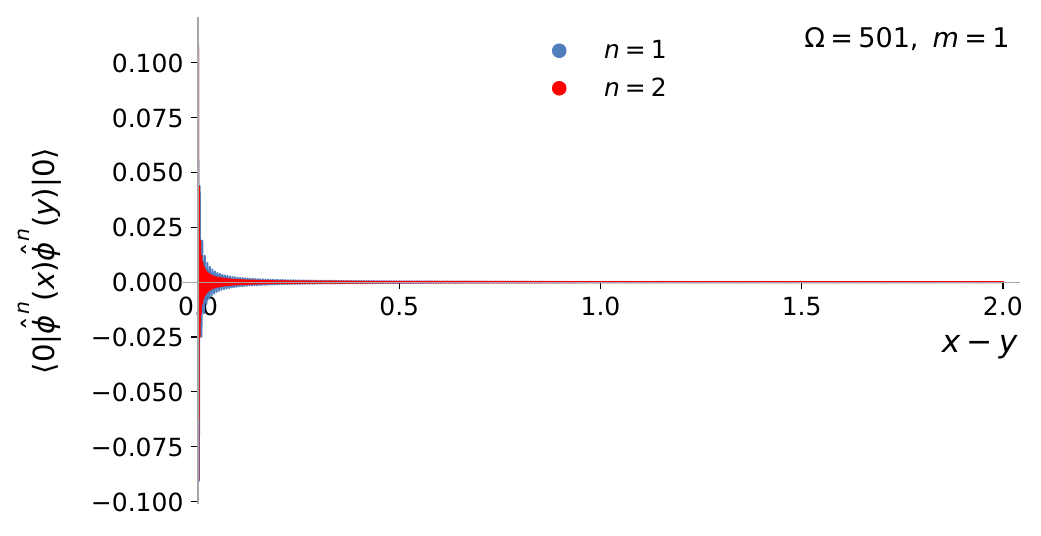}
    \caption{Plots of the equal-time correlation functions for the wavelet components $\hat{\phi}^n$ as functions of $x-y$ for $\Omega=501$, $m=1$, with the same band-field normalisation as Fig.~\ref{fig:waveletcorrelationfunction}. The blue plot is at $n=1$ and the red plot is at scale $n=2$. Note that the horizontal axis runs from 0 to 2.0. The rapid oscillations reflect the shorter length scale $1/a_n$; the correlations still have oscillatory tails at larger separations.}
    \label{fig:largeratiowaveletcorrelationfunction}
\end{figure}

From Eq.~\eqref{Hamiltoniandecomposition}, we can see that in the Fourier domain, the Hamiltonian for the Klein--Gordon field can be written as
\begin{align}
    \hat{H}_{KG} 
    =&
    \sum_{n\in\integers} \hat{H}^n
\\
    =&
    \int_\mathbb{R}\frac{dk}{2\pi}\omega_k\hat{a}_k^\dagger \hat{a}_k \Bigg(
    \nonumber
    \sum_{n\in\integers}\chi_n(k)\Bigg).
\end{align}
Now for the components of $\hat{\phi}$ at each scale, given by
\begin{align}
    \hat{\phi}^n(x)
    &= 
    \int_\mathbb{R}\frac{dk}{2\pi}\hat{\tilde{\phi}}(k)\chi_n(k)e^{ikx},
\end{align}
the following is true
\begin{subequations}
\begin{align}
    e^{i\hat{H}t}\hat{\phi}^n(x)e^{-i\hat{H}t} 
    &= 
    e^{i\hat{H}^n t}\hat{\phi}^n(x)e^{-i\hat{H}^n t}.
\end{align}
\label{decoupled time evolution}
\end{subequations}
This implies that the time evolution of the wavelet components under the total Hamiltonian is equivalent to the Heisenberg evolution under the corresponding scale component of the Hamiltonian. In fact, we can write
\begin{align}
    e^{i\hat{H}^nt}\hat{\phi}^m(x)e^{-i\hat{H}^nt}= \hat{\phi}^m(x, \delta^{nm} t).
\end{align}
That being said, quantities such as the ground state two-point correlation functions for the scale components of $\hat{\phi}$ will be zero for different scales, and will be equivalent to the twopoint correlation functions for a bandlimited field with UV and IR cut-offs. We have
\begin{align}
    \bra{0}\hat{\phi}^n(x, t)\hat{\phi}^m(y, t')\ket{0} = \delta^{nm}\bra{0}\hat{\phi}^n(x, t)\hat{\phi}^n(y, t')\ket{0}.
\end{align}
From the previous section, we know what such a correlation function is given by

\begin{align}
    &\bra{0}\hat{\phi}^n(x, t)\hat{\phi}^n(y, t')\ket{0}
    \nonumber\\*
    &\quad=
    \int_{-2^{n+1}\Omega}^{-2^n\Omega}\frac{dk}{2\pi}\frac{1}{2\omega_k}e^{i(k(x-y)-\omega_k(t-t'))} 
    \nonumber
    \\*
    &\qquad+ \int_{2^{n}\Omega}^{2^{n+1}\Omega}\frac{dk}{2\pi}\frac{1}{2\omega_k}e^{i(k(x-y)-\omega_k(t-t'))}.
\end{align}
For equal-time separations, this is given by
\begin{align}
    \bra{0}\hat{\phi}^n(x)\hat{\phi}^n(y)\ket{0}
    &=\int_\reals\frac{dk}{2\pi}\chi_n(k)
        \frac{\cos(k(x-y))}{2\sqrt{k^2+m^2}}
    \nonumber\\
    &=\int_{a_n}^{2a_n}\frac{dk}{2\pi}
        \frac{\cos(k(x-y))}{\sqrt{k^2+m^2}}.
    \label{equaltimebandcorrelator}
\end{align}
A plot of this correlation function for $\Omega=m=1$ for two different scales is shown in Fig.~\ref{fig:waveletcorrelationfunction}.
When the band momenta are much greater than the mass of the field, the characteristic oscillation length is set by $1/a_n$. Increasing the band scale therefore gives oscillations over shorter physical separations, as illustrated in Fig.~\ref{fig:largeratiowaveletcorrelationfunction}. The sharp band edges also produce long oscillatory tails, so this length should not be interpreted as a finite range beyond which the correlation is zero.

So, at each scale, we have correlation functions of the form shown in Fig.~\ref{fig:waveletcorrelationfunction}, while the corresponding lattice spacing is $h_n=\pi/a_n$. To compare the physics at different scales, it is useful to express the separation in units of this spacing. Setting $r=(x-y)/h_n$ and $s=k/a_n$ in Eq.~\eqref{equaltimebandcorrelator} gives
\begin{align}
    \bra{0}\hat{\phi}^n(x)\hat{\phi}^n(y)\ket{0}
       &=\int_1^2\frac{ds}{2\pi}
          \frac{\cos(\pi sr)}{\sqrt{s^2+(m/a_n)^2}}.
\end{align}
We can see here that at fixed non-zero mass the dimensionless ratio $m/a_n$ still depends on scale. Consequently, expressing distances in lattice units does not make the massive theories at different scales identical. The rescaled band correlators coincide for a massless field, and approach that form for bands with $a_n\gg m$. For the canonical coefficients $Q^n_j$, the same comparison also includes the factor $h_n$ from Eq.~\eqref{quantumsamplingdictionary}.

\section{Dyadic co-ordinates with Shannon wavelets and quantum fields \label{sec:dyadicrepfreescalarfield}}

\subsection{Sampling theory on the dyadic rational co-ordinates}
We now relabel the orthonormal wavelets already defined in Eq.~\eqref{orthonormalshannonwavelet}; their normalization does not change. The half-site sampling position defines
\begin{align}
    \eta&\coloneqq\frac{\Omega x_{nj}}{\pi}=2^{-n}(j+1/2),&
    u(x;\eta)&\coloneqq u^n_j(x).
    \label{normshiftshannonwavdef}
\end{align}
Writing $\mathbb D\coloneqq\mathbb Z[1/2]$ for the dyadic rationals, the numbers with finite binary expansions, the wavelet sites form $\mathbb D^\times=\mathbb D\setminus\{0\}$. The set $\mathbb{D}^\times$ is defined to consist of the non-zero numbers with finite binary expansions~\cite{LynchMackey2024,Daubechies1992,AntoniadisGregoireMcKeague1994}. This is given explicitly by 
\begin{align}
    \mathbb{D}^\times \coloneqq&~ \integers [1/2]\setminus\{0\}
    \nonumber
    \\
    =&~ \left\{\frac{m}{2^r}~:~m\in\integers,~r\in\integers \right\}\setminus\{0\}.
\end{align}
The strings on either side of the binary point may be arbitrarily long, but the position of the rightmost `$1$' determines its dyadic valuation $\nu_2(\eta)$. For example, consider $\eta = (110.01)_2 = 25/4$. This number ends with a $1$ in the `quarters' place so its dyadic valuation is $-2$. By comparison, $(110.00)_2 =6$ has its rightmost $1$ in the twos place, so its valuation is $1$. In each case, the place value of that bit is the power of two that can be factored out to leave an odd integer. Zero has no such bit and so we exclude it from the set and from the set of available lattice sites for the wavelets. Additionally, no finite $(n,j)$ maps to zero. Table~\ref{tab:dyadicexamples} compares the two examples. Note that despite having a comparitively similar magnitude in terms of real valuation, the dyadic magnitude of 6.25 is eight times the dyadic magnitude of 6.

\begin{table}[!h]
    \caption{The dyadic magnitude $|\eta|_2=2^{-\nu_2(\eta)}$ of $25/4$ is eight times that of $6$, although their decimal values are close.}
    \label{tab:dyadicexamples}
    \begin{ruledtabular}
        \begin{tabular}{cccc}
            Decimal & Binary & \shortstack{Dyadic\\valuation $\nu_2(\eta)$}
                & \shortstack{Dyadic\\magnitude $|\eta|_2$} \\
            \hline
            $6.25$ & $(110.01)_2$ & $-2$ & $4$ \\
            $6$ & $(110.00)_2$ & $1$ & $1/2$
        \end{tabular}
    \end{ruledtabular}
\end{table}

\begin{figure}[t]
    \centering
    \includegraphics[width=\linewidth]{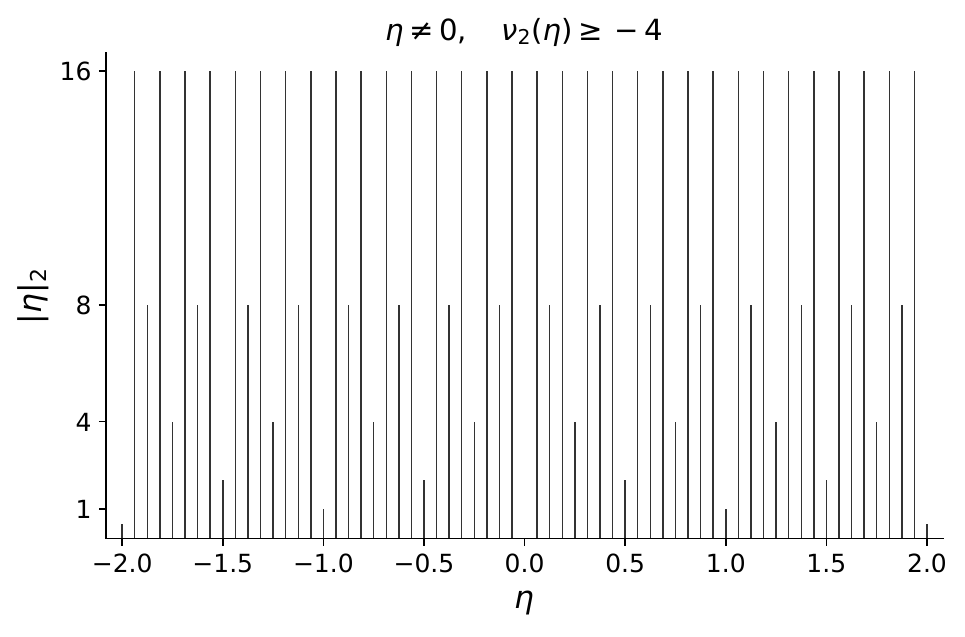}
    \caption{Representation of the lattice of dyadic rational co-ordinates plotted between $-2$ and $2$, with height $|\eta|_2$. Note that only $\eta$ with dyadic valuation $\nu_2(\eta)\geq-4$ ($|\eta|_2\leq16$) are present here to maintain visual clarity. The origin $\eta=0$ is excluded, as it does not represent a wavelet site in $\mathbb D^\times$.}
    \label{fig:dyadiccoordinates}
\end{figure}

More formally, for $\eta\neq 0$ the dyadic valuation $\nu_2(\eta)$ is the largest integer $k$ for which $2^{-k}\eta$ is an integer. We have
\begin{align}
    \nu_{2}(\eta) \coloneqq 
    \begin{cases}
        \infty &\text{if } \eta = 0\\
        \max{\{k\in{\integers}:2^k\mid\eta\}} &\text{if }\eta \neq 0.
    \end{cases}
\end{align}  
Note that for $\eta\in\mathbb{D}^\times$ there is no $\eta=0$ and the first line of the definition of the dyadic valuation is unneeded for the wavelet lattice sites. We have included it in its definition for consistency with existing literature.  
The inverse map between the dyadic rational and $(n,j)$ representations 
\begin{subequations}
\begin{align}
    n&=-\nu_2(\eta)-1,&
    j&=2^n\eta-\frac12,
\end{align}
\label{dyadiclatticecoords}
\end{subequations}
where $\nu_2(\eta)$ is the integer exponent of two in the unique expression $\eta=2^{\nu_2(\eta)}r$ with $r$ odd. The dyadic magnitude is
\begin{align}
    |\eta|_2&\coloneqq 2^{-\nu_2(\eta)}=2^{n+1},&
    h_\eta&\coloneqq\frac{2\pi}{|\eta|_2\Omega}=h_n,
    \nonumber\\
    x_\eta&\coloneqq\frac{\pi}{\Omega}\eta=x_{nj}.
    \label{dyadicnormalizationdictionary}
\end{align}
The lattice sites and their dyadic magnitudes are illustrated in Fig.~\ref{fig:dyadiccoordinates}. This dyadic structure is standard in multiresolution and wavelet analysis \cite{Chui1992,Daubechies1992,Mallat2008}; the valuation is familiar from $p$-adic analysis \cite{Gouvea1997,Koblitz1984}.

In these coordinates the same orthonormal wavelet is
\begin{subequations}
\begin{align}
    u(x;\eta)&=\frac{1}{\sqrt{h_\eta}}
       \Bigl[2\sinc\bigl(|\eta|_2\Omega(x-x_\eta)\bigr)
       \nonumber\\
       &\hspace{28mm}-\sinc\bigl(|\eta|_2\Omega(x-x_\eta)/2\bigr)\Bigr],\\
    \widetilde u(k;\eta)&=\sqrt{h_\eta}\,e^{-ikx_\eta}\chi_\eta(k).
    \label{dyadicwaveletfourier}
\end{align}
\end{subequations}
The dyadic band indicator is the indicator of Eq.~\eqref{waveletbandindicator} with its scale relabelled:
\begin{align}
    \chi_\eta(k)&\coloneqq
      \mathbf{1}_{\{|\eta|_2\Omega/2<|k|<|\eta|_2\Omega\}}
      \nonumber\\
      &=\Pi\left(\frac{k}{2|\eta|_2\Omega}\right)
        -\Pi\left(\frac{k}{|\eta|_2\Omega}\right).
    \label{dyadicrationalindicatormomentum}
\end{align}
The top hat expression again holds almost everywhere with the endpoint convention of Sec.~\ref{sec:waveletdecomposition}. The indicator depends on $\eta$ only through $|\eta|_2$, whereas the wavelet also depends on the position $x_\eta$.

Orthonormality becomes
\begin{align}
    \int_\reals dx\,\overline{u(x;\eta)}u(x;\eta')
       &=\delta[\eta-\eta'],
\end{align}
where the square brackets denote a Kronecker delta on the wavelet%
\footnote{We use the square bracket notation here to differentiate the Kronecker delta on a countably dense set such as the dyadic rationals $\mathbb{D}^{\times}$ or the real numbers from the usual Dirac delta prescription usually defined with parentheses as notation with $\delta(x)$.} %
 sites with
\begin{align}
    \delta[\eta-\eta'] \coloneqq 
    \begin{cases}
    1, &\eta=\eta',\\
    0, &\text{otherwise.}
    \end{cases}
\end{align} 

The function expansion and coefficient dictionary are
\begin{align}
    f(x)&=\sum_{\eta\in\mathbb D^\times}F_\eta u(x;\eta),\\
    F_\eta&\coloneqq\int_\reals dx\,\overline{u(x;\eta)}f(x)
       =F^n_j=\sqrt{h_\eta}\,f^n(x_\eta).
    \label{dyadiccoefficientdictionary}
\end{align}
Thus $F_\eta$ is an orthonormal coefficient; the literal sample is $F_\eta/\sqrt{h_\eta}$.

The inner product contains no sampling weight in these coefficients:
\begin{align}
    \int_\reals dx\,\overline{f(x)}g(x)
       &=\sum_{\eta\in\mathbb D^\times}\overline{F_\eta}G_\eta
        =\sum_{n,j}\overline{F^n_j}G^n_j.
    \label{dyadicrationalinnerprod}
\end{align}
This is precisely Eq.~\eqref{shannonwaveletinnerprod} after applying the coefficient dictionary in Eq.~\eqref{dyadiccoefficientdictionary}.

Likewise, the derivative matrix is the same operator with relabelled indices:
\begin{align}
    D^{(\ell)}_{\eta\eta'}
       &\coloneqq\int_\reals dx\,\overline{u(x;\eta)}
                       \partial_x^\ell u(x;\eta')
       =D^{(\ell)nm}_{jh}
       \nonumber\\
       &=\sqrt{h_\eta h_{\eta'}}\int_\reals\frac{dk}{2\pi}(ik)^\ell
          e^{ik(x_\eta-x_{\eta'})}\chi_\eta(k)\chi_{\eta'}(k),
    \label{dyadicderivativematrix}
\end{align}
where $\eta$ corresponds to $(n,j)$ and $\eta'$ to $(m,h)$. Its action on the coefficients $F_\eta$ and its reinterpolation via the wavelets are given by
\begin{subequations}
\begin{align}
    (\partial^\ell F)_\eta
       &=\sum_{\eta'\in\mathbb D^\times}D^{(\ell)}_{\eta\eta'}F_{\eta'},\\
    \partial_x^\ell f(x)
       &=\sum_{\eta,\eta'\in\mathbb D^\times}
           u(x;\eta)D^{(\ell)}_{\eta\eta'}F_{\eta'}.
\end{align}
\end{subequations}
Because $\chi_\eta\chi_{\eta'}=0$ for $|\eta|_2\neq|\eta'|_2$, the derivative couples only equal dyadic magnitudes, exactly as it couples only equal scales in the $(n,j)$ representation.

\subsection{The free scalar field on a dyadic lattice}
Performing a Shannon wavelet decomposition using a dyadic co-ordinate representation gives some insight as to how the different wavelet scales couple together. 

Starting with the usual operator valued scalar function satisfying the Klein--Gordon equation, we have
\begin{align}
    \square \op{\phi}(x) + m^2\op{\phi} = 0.
\end{align}
Treating the field in its usual creation and annihilation operator representation, we can decompose it into a Shannon wavelet basis on a lattice of dyadic rationals.
\begin{align}
    \op{\phi}(x) =& \int_\reals \frac{dk}{2\pi} \frac{1}{\sqrt{2\omega_k}}\left(\op{a}_ke^{ikx} + \op{a}_k^\dagger e^{-ikx}\right)
    \\*
    \nonumber
    =&
    \sum_{\eta\in\mathbb D^\times}\op{Q}_\eta u(x;\eta),
\end{align}
where $\op{Q}_\eta=\op{Q}^n_j$ is the orthonormal field coefficient at the site $\eta$, given by
\begin{align}
    \op{Q}_\eta \coloneqq &\int_\reals dx u^*(x;\eta)\op{\phi}(x)
    \\
    \nonumber
    =& 
    \int_\reals \frac{dk}{2\pi} \overline{\widetilde u(k;\eta)} \op{\tilde{\phi}}(k).
\end{align}
Likewise $\op{P}_\eta=\op{P}^n_j$ is the coefficient of the conjugate momentum field, with $\op{Q}_\eta=\sqrt{h_\eta}\op{\phi}^{n}(x_\eta)$ and $\op{P}_\eta=\sqrt{h_\eta}\op{\pi}^{n}(x_\eta)$. We have
\begin{align}
    \op{\pi}\coloneqq& \int_\reals \frac{dk}{2\pi}(-i)\sqrt{\frac{\omega_k}{2}}\left(\op{a}_k e^{ikx} -\op{a}_{k}^\dagger e^{-ikx}\right)
    \nonumber
    \\
    =& 
    \sum_{\eta\in\mathbb D^\times}\op{P}_\eta u(x;\eta).
\end{align}
The creation and annihilation operators for both the field and its conjugate satisfy the usual Dirac delta commutation relations for a scalar field theory. However, the commutation relations for the coefficients of the field and its conjugate on the dyadic rational lattice depend on the orthogonality of the Shannon wavelets. In that we have
\begin{subequations}
\begin{align}
    \left[\op{\phi}(x),\op{\pi}(y)\right] =& i\delta (x-y)
    \\
    \left[\op{Q}_\eta,\op{P}_{\eta'}\right] =& i\delta[\eta - \eta'],
\end{align}
\end{subequations}
where again, the parentheses on the delta function indicate the usual Dirac delta function conditions while the square brackets on a delta function indicate Kronecker delta behaviour on the dyadic co-ordinates. All other field commutators are zero. 

Relabelling Eq.~\eqref{waveletHamiltonian} gives the same free Hamiltonian:
\begin{align}
    \op H_{KG}
       &=\frac12\sum_{\eta\in\mathbb D^\times}
           \left[\op P_\eta^2+m^2\op Q_\eta^2\right]
       \nonumber\\*
       &\quad-\frac12\sum_{\eta,\eta'\in\mathbb D^\times}
           \op Q_\eta D^{(2)}_{\eta\eta'}\op Q_{\eta'}.
    \label{dyadiccanonicalhamiltonian}
\end{align}
Here $D^{(2)}_{\eta\eta'}$ is the second-derivative matrix defined in Eq.~\eqref{dyadicderivativematrix}. It couples only sites with equal dyadic magnitude. The Hamiltonian and canonical variables therefore agree with the $(n,j)$ representation without any additional normalization factors.

\begin{figure}[t]
    \centering
    \includegraphics[width=\linewidth]{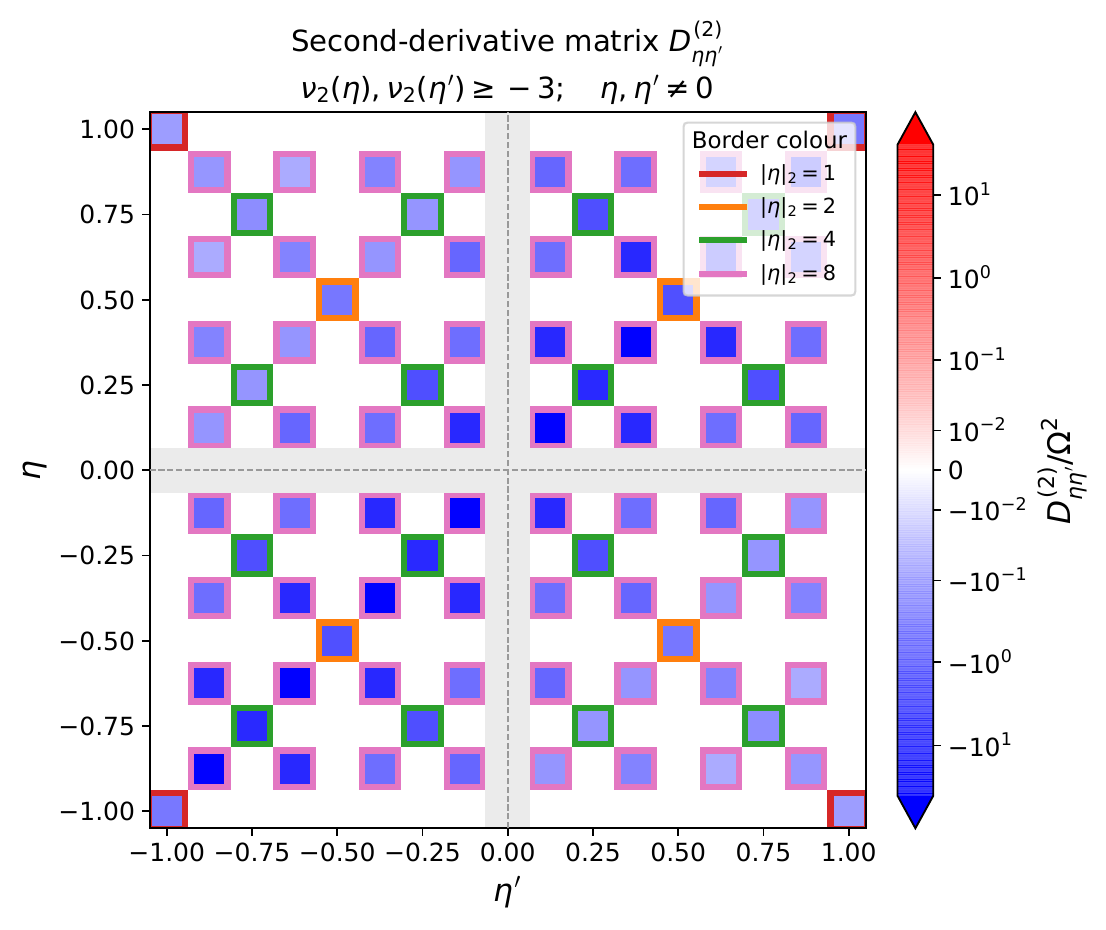}
    \caption{Plot of the dimensionless matrix elements $D^{(2)}_{\eta\eta'}/\Omega^2$ for the dyadic second derivative on a lattice of dyadic rationals. For clarity, the sites lie between $-1$ and $1$, with $\nu_2(\eta),\nu_2(\eta')\geq-3$ and the origin excluded. It is clear here that when the dyadic magnitudes of $\eta$ and $\eta'$ are unequal, the matrix element is zero. The lattice sites are also given coloured borders to indicate their common dyadic magnitude, showing how it affects the coupling strength for the second derivative wave operator. The fill is blue for negative entries, with opacity increasing logarithmically with magnitude.}
    \label{fig:dyadicderiv2}
\end{figure}

The elements of this matrix are given by
\begin{align}
    D^{(2)}_{\eta\eta'} =
    \begin{cases}
        \frac{-7\Omega^2|\eta|_2^2}{12}, 
        &\eta=\eta'
        \\
        \frac{2\Omega^2 \delta[|\eta|_2-|\eta'|_2]}{\pi^2}\left(\frac{(-1)^{\frac{|\eta|_2}{2}(\eta-\eta')}-2}{(\eta-\eta')^2}\right), 
        &\eta\neq\eta'.
    \end{cases}
    \label{dyadicwaveoperator}
\end{align}
We can see here that due to the $\delta[|\eta|_2-|\eta'|_2]$, only samples of the field at co-ordinates with the same dyadic norm are coupled together. A plot of these derivative matrix elements is given in Fig.~\ref{fig:dyadicderiv2}. 
For joint lattice sites with the same dyadic magnitude $|\eta|_2$ as seen along the vertical, horizontal, or same border colour lines in Fig.~\ref{fig:dyadicderiv2}, we see the polynomial decay in the strength of the coupling between these lattice sites along with the oscillation given by the numerator of Eq.~\eqref{dyadicwaveoperator}. Note here that the elements of this matrix $[D^{(2)}]_{\eta\eta'}$ are strictly negative. This negativity is parallel to the negativity of the matrix elements of the second derivatives of the Shannon wavelets themselves on the regular lattice sites. 

As the free scalar field operators at any given lattice site are only coupled to those at lattice sites with the same dyadic norm, we show now that it is only those coupled oscillators that contribute to the time evolution of any one point of the field via the usual Heisenberg time evolution
\begin{align}
    \op{\phi}(x, t) = e^{i\op{H}t}\op{\phi}(x, 0)e^{-i\op{H}t}.
\end{align}
The canonical commutators and Eq.~\eqref{dyadiccanonicalhamiltonian} give
\begin{align}
    [\op Q_\eta,\op H_{KG}]&=i\op P_\eta,\\
    [\op P_\eta,\op H_{KG}]&=
        i\sum_{\eta'}D^{(2)}_{\eta\eta'}\op Q_{\eta'}-im^2\op Q_\eta.
\end{align}
Thus $\partial_t\op Q_\eta=\op P_\eta$ and
$\partial_t\op P_\eta=\sum_{\eta'}D^{(2)}_{\eta\eta'}\op Q_{\eta'}-m^2\op Q_\eta$, exactly as in the canonical $(n,j)$ representation.

\begin{figure}[t]
    \centering
    \includegraphics[width=\linewidth]{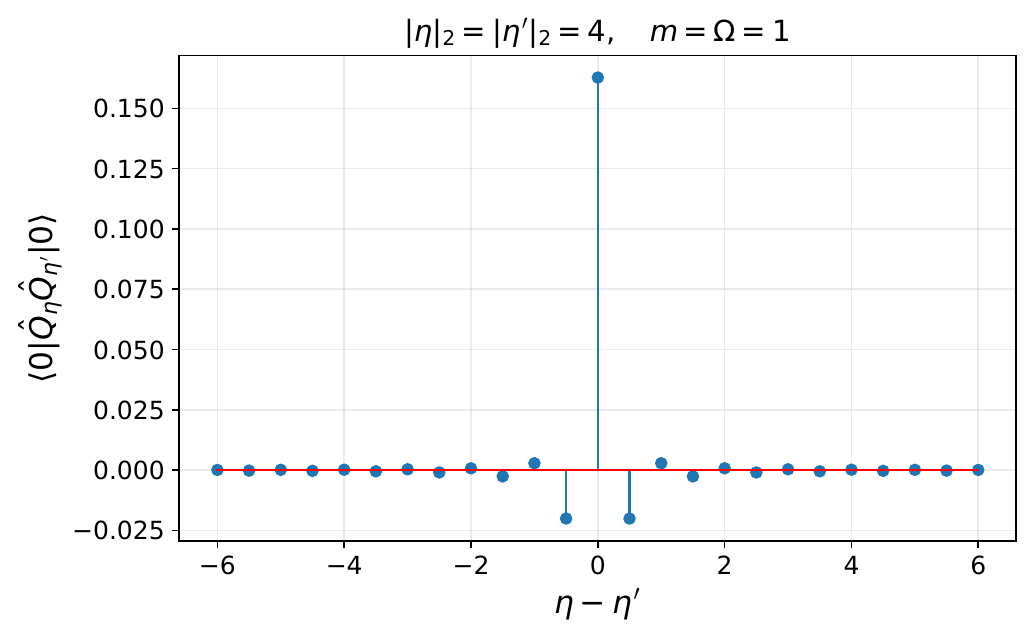}
\caption{Plot of the equal-time ($t=t'$) two-point correlation function $\bra0\op Q_\eta(t)\op Q_{\eta'}(t)\ket0$ for dyadic separation $\eta-\eta'$ on the $\mathbb D^\times$ lattice from Eq.~\eqref{dyadiccorrelationfunction}. These are canonical orthonormal coefficients rather than unnormalised field samples. Note that non-zero correlations occur only for equal dyadic magnitudes $|\eta|_2=|\eta'|_2$. Here, $|\eta|_2=|\eta'|_2=4$ and the other parameters have been set to $m=\Omega=1$.}
    \label{fig:dyadictwopoint}
\end{figure}

For a fixed dyadic magnitude $q$, define its sector Hamiltonian by
\begin{align}
    \op H_{KG}|_q&\coloneqq
       \frac12\sum_{\substack{\xi\in\mathbb D^\times\\|\xi|_2=q}}
          \left[\op P_\xi^2+m^2\op Q_\xi^2\right]
       \nonumber\\
       &\quad-\frac12
        \sum_{\substack{\xi,\xi'\in\mathbb D^\times\\|\xi|_2=|\xi'|_2=q}}
           \op Q_\xi D^{(2)}_{\xi\xi'}\op Q_{\xi'}.
\end{align}
For $q=|\eta|_2$, this is $\op H^n$ with $q=2^{n+1}$. Therefore
\begin{subequations}
\begin{align}
    \op Q_\eta(t)
       &=e^{i\op H_{KG}t}\op Q_\eta e^{-i\op H_{KG}t}
       \nonumber\\
       &=e^{i(\op H_{KG}|_q)t}\op Q_\eta e^{-i(\op H_{KG}|_q)t},\\
    \op P_\eta(t)
       &=e^{i\op H_{KG}t}\op P_\eta e^{-i\op H_{KG}t}
       \nonumber\\
       &=e^{i(\op H_{KG}|_q)t}\op P_\eta e^{-i(\op H_{KG}|_q)t}.
\end{align}
\end{subequations}
It follows then that functions such as the two-point correlation function between two dyadic rational co-ordinates should only be non-zero for co-ordinates of the same dyadic norm. In fact, we can see this is true by direct evaluation of the two point function for the free field. We have 
\begin{align}
    \bra{0}\op{Q}_\eta(t)&\op{Q}_{\eta'}(t')\ket{0}
    \nonumber
    \\
    =&\sqrt{h_\eta h_{\eta'}}\int_\reals \frac{dk}{2\pi}\frac{1}{2\omega_k}
    \nonumber\\
    &\times e^{ik(x_\eta-x_{\eta'})-i\omega_k(t-t')}\chi_\eta(k)\chi_{\eta'}(k).
\end{align}
Now, given that for any two $\eta,\eta'$ with differing dyadic magnitudes, $\chi_{\eta}\chi_{\eta'} = 0$, we can write
\begin{align}
    \chi_{\eta}(k)\chi_{\eta'}(k) = \delta[|\eta|_2-|\eta'|_2]\chi_{\eta}(k).
\end{align}
Thus, the two-point correlator between two orthonormal field coefficients can be written as
\begin{align}
    \bra{0}\op{Q}_\eta(t)&\op{Q}_{\eta'}(t')\ket{0}
    \nonumber
    \\
    =&h_\eta\delta[|\eta|_2-|\eta'|_2]\int_\reals \frac{dk}{2\pi}\frac{1}{2\omega_k}
    \nonumber\\
    &\times e^{ik(x_\eta-x_{\eta'})-i\omega_k(t-t')}\chi_\eta(k),
    \label{dyadiccorrelationfunction}
\end{align}
where the only vacuum correlations between any two samples of the field are those between co-ordinates of the same dyadic norm. This is analogous to the earlier result that only field samples on the same scale layer $n$ have non-zero vacuum correlations.
A plot of the equal time two-point correlation function between orthonormal field coefficients on the dyadic rational lattice is given in Fig.~\ref{fig:dyadictwopoint}.
Note that this has been presented as discrete in $\eta-\eta'$ due to the fact that they are only non-zero only at positions when $|\eta|_2 = |\eta'|_2$.

\section{Shannon wavelets and interacting scalar fields \label{sec:interactingwaveletfield}}

Given that a free scalar field decomposes entirely into independent wavelet scales as seen in both the two-point correlation functions and in Eqs.~\eqref{Hamiltoniandecomposition} and~\eqref{decoupled time evolution}, it is useful to study the wavelet decompositions of Hamiltonians of order higher than quadratic. We can expect such Hamiltonians to couple different scales because orthogonality of two wavelets does not require the overlap
\begin{align}
    \int_\reals dx\,u^n_j(x)\left(u^m_k(x)\right)^\ell
\end{align}
to vanish when $m\neq n$ and $\ell\geq2$. Some choices of scales and translations still give zero, as the support conditions below will make clear. Now, given that fields that self interact do not fully decouple under a wavelet decomposition, it would be prudent to study the nature (and effective range) of the interactive coupling between wavelet scales (using a %
$\op{\phi}^4$ term) as well as correlations between these scales.
\subsection{Sampling of quartic Hamiltonians using wavelets}
In order to construct a scalar field that can self interact we need to include a term to its Hamiltonian that is a polynomial of the field of order greater than two. Here we focus on a stable quartic interaction with $\lambda>0$ as a simple model for interactions in quantum field theory. 

The usual equation of motion for an interacting scalar field on Minkowski spacetime is given by
\begin{align}
    \square \op{\phi} +m^2\op{\phi} = -\frac{\lambda}{3!}\op{\phi}^3.
\end{align}
The Hamiltonian for this field in 1+1 dimensions can be split into its free and interacting components as
\begin{align}
    \op H&=\op H_{KG}+\op H_I,&
    \op H_I&\coloneqq\frac{\lambda}{4!}\int_\reals dx\,\op\phi(x)^4.
    \label{quarticinteractionconvention}
\end{align}
Here $\op H_{KG}$ is the free Hamiltonian of Eq.~\eqref{KGHam2}, including the factor $1/2$ in its quadratic terms. We use $\op H_I$ for the complete quartic interaction, with the coefficient $\lambda/4!$ included, throughout both lattice representations. 

The free component's treatment in the wavelet basis is none other than what was presented in the previous section, with all components of the field decoupling into their respective isolated scale components as seen in Eq.~\eqref{discreteHamiltonians} and Eq.~\eqref{Hamiltoniandecomposition}. The interacting component is not so simple. Taking just the interacting component and rewriting the individual fields in terms of their wavelet basis decomposition, we have
\begin{align}
    \op H_I
&=\frac{\lambda}{4!}
    \int_\reals dx\left(\op{\phi}(x)\right)^4
    \nonumber
\\
&=\frac{\lambda}{4!}
    \int_\reals dx\left[\left(\sum_{n\in\integers}\sum_{j\in\integers}\op{Q}^n_ju^n_j(x)\right)^4\right].
\end{align}
This can be written explicitly on a wavelet-based sampling lattice as
\begin{align}
    \op H_I = \frac{\lambda}{4!}\sum_{\bvec{n} \in\integers^4}\sum_{\bvec{j}\in\integers^4} \op{Q}^{n_1}_{j_1}\op{Q}^{n_2}_{j_2}\op{Q}^{n_3}_{j_3}\op{Q}^{n_4}_{j_4} \mathcal{E}^{n_1,n_2,n_3,n_4}_{j_1,j_2,j_3,j_4},
\end{align}
where $\mathcal E$ is the overlap tensor of the orthonormal wavelets throughout both lattice representations:
\begin{align}
    \mathcal{E}^{n_1,n_2,n_3,n_4}_{j_1,j_2,j_3,j_4} \coloneqq \int_\reals dx u^{n_1}_{j_1}(x)u^{n_2}_{j_2}(x)u^{n_3}_{j_3}(x)u^{n_4}_{j_4}(x).
\end{align}
Note that under permutation of the labels for these indices, the value of $\mathcal{E}$ is left entirely unchanged. We can evaluate this by considering the Fourier representations of the wavelets and evaluating the integral over $x$ as
\begin{align}
    \mathcal{E}_{j_1,j_2,j_3,j_4}^{n_1,n_2,n_3,n_4} =&
    \sqrt{h_{n_1}h_{n_2}h_{n_3}h_{n_4}}\int_\reals \frac{dk_1dk_2dk_3}{(2\pi)^3}e^{-i\Phi(\bvec{k})}
    \nonumber
    \\
    &\times
    \prod_{r=1}^3\chi_{n_r}(k_r)\chi_{n_4}(-k_1-k_2-k_3),
    \label{quarticoperatorfourierrep}
\end{align}
where $\Phi(\bvec{k})$ is given by
\begin{align}
    \Phi(\bvec{k}) \coloneqq &
    k_1(x_{n_1j_1}-x_{n_4j_4})
    \nonumber
    \\
    &+ k_2(x_{n_2j_2}-x_{n_4j_4}) + k_3(x_{n_3j_3}- x_{n_4j_4}),
\end{align}
and $\chi_n$ is the band indicator already defined in Eq.~\eqref{waveletbandindicator}, with support $a_n<|k|<2a_n$. Each factor $\sqrt{h_{n_r}}$ in Eq.~\eqref{quarticoperatorfourierrep} comes from one orthonormal Fourier wavelet.

The product over indicator functions in this manner and the elimination of the integral over $k_4$ arises due to the fact that each of the wavelets contain a complex exponential in $x$ and $k_i$ and
\begin{align}
    \int_\reals dx e^{ix(k_1 + k_2 + k_3 + k_4)} = 2\pi\delta(k_1 + k_2 + k_3 + k_4),
    \label{momentumconservationfourier}
\end{align}
which can be interpreted as a requirement for the conservation of momentum in the field theory. 
It is useful to note now that a common translation for all four wavelets in scale will only change $\mathcal{E}$ by a global factor, in that be have
\begin{align}
    \mathcal{E}^{n_1+a, n_2+a,n_3+a,n_4+a}_{j_1,j_2,j_3,j_4} = 2^{a}\mathcal{E}_{j_1,j_2,j_3,j_4}^{n_1,n_2,n_3,n_4}.
\end{align}
We can also see that the product of indicator functions in Eq.~\eqref{quarticoperatorfourierrep}, together with Eq.~\eqref{momentumconservationfourier}, gives a condition on which sets of scales can contribute to the quartic coupling tensor. For four momenta within their respective bands, let $S$ be the non-empty set of labels with positive momentum and $S^c$ its non-empty complement with negative momentum. Their total magnitudes lie in the intervals
\begin{subequations}
    \begin{align}
        \Omega\sum_{q\in S}2^{n_q}
           &<\sum_{q\in S}|k_q|
             <2\Omega\sum_{q\in S}2^{n_q},\\
        \Omega\sum_{p\in S^c}2^{n_p}
           &<\sum_{p\in S^c}|k_p|
             <2\Omega\sum_{p\in S^c}2^{n_p}.
    \end{align}
\end{subequations}
Momentum conservation is possible within these bands precisely when the two open intervals overlap for at least one such partition. Thus the support condition is that there exists a choice of $S$ for which
\begin{align}
    \frac{1}{2}<\frac{\sum_{q\in S}2^{n_q}}{\sum_{p\in S^c}2^{n_p}}<2.
    \label{interactionadmissability}
\end{align}
It is important to distinguish this condition on the momentum support from the value of a particular coupling coefficient. If no partition satisfies Eq.~\eqref{interactionadmissability}, then $\mathcal E$ vanishes for every choice of translations $j_i$. If the condition is satisfied, momentum conservation is kinematically possible, but the translation-dependent phase in Eq.~\eqref{quarticoperatorfourierrep} may still cause the integral to vanish. The condition is therefore necessary, but not sufficient, for a specified coefficient to be non-zero.

For example, the choice of scales $(0,0,0,3)$ gives $\mathcal E^{0003}_{j_1,j_2,j_3,j_4}=0$ for every choice of translations, whereas $(0,0,0,2)$ is support-admissible. A direct example of cancellation within admissible support is supplied by Eq.~\eqref{couplingtensor3-1slice}: four equal scales satisfy the support condition, yet $\mathcal E^{nnnn}_{j,j,j,j+d}=0$ for odd $d$. Meanwhile, paired scales $(n,n,m,m)$ always admit momentum conservation, since one can balance opposite momenta within each pair. Thus support alone does not impose a maximum separation between two paired scales. These are statements about spatial momentum support of the overlap tensor; scattering amplitudes would additionally require the appropriate on-shell and energy-conservation conditions.

\begin{figure}[t]
    \centering
    \includegraphics[width=\linewidth]{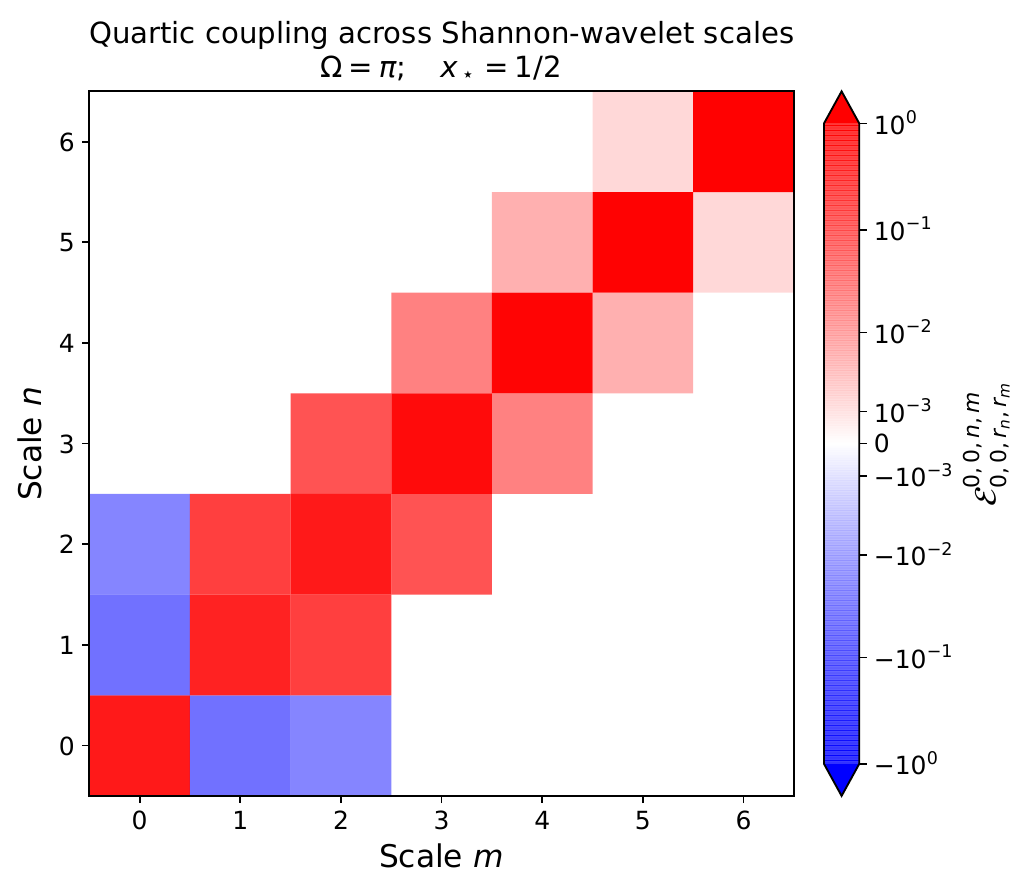}
    \caption{Plot of the quartic coupling strength $\mathcal E^{00nm}_{00r_nr_m}$ for varying scales $n,m=0,\ldots,6$, with $\Omega=\pi$ and the local sites $r_n,r_m$ chosen nearest to $x_\star=1/2$. The tensor is formed from orthonormal wavelets and does not include the interaction prefactor $\lambda/4!$. The paired slice $\mathcal E^{00mm}_{00r_mr_m}$ remains positive at arbitrary scale separation. Other coefficients vanish whenever their scales fail Eq.~\eqref{interactionadmissability}; satisfying that condition alone does not guarantee a non-zero value at specified translations. Here and in Figs.~\ref{fig:interactioncouplingplotjsep}--\ref{fig:quartictensorsepr}, blue and red indicate negative and positive values, respectively, with opacity increasing logarithmically with magnitude.}
    \label{fig:interactioncouplingplotnsep}
\end{figure}

A plot of this is shown in Fig.~\ref{fig:interactioncouplingplotnsep}.

In order to analyse the structure of this eight index coupling tensor it is useful to define the local lattice parameter $r_n$ such that the lattice site $x_{nr_n}$ is as close to $1/2$ as can be reached for a given scale $n$. We have
\begin{align}
    r_n \coloneqq
    \begin{cases}
        0, &n=0
        \\
        2^{n-1}-1, &\text{otherwise}
    \end{cases}
    .
\end{align}
The lattice position $x_{nj_n}$ is thus given by
\begin{align}
    x_{nr_n} =& 2^{-n}\frac{\pi}{\Omega}(r_n + 1/2)
    \nonumber
    \\
    =&
    \begin{cases}
        \frac{\pi}{\Omega}\frac{1}{2}, &n=0,
        \\
        \frac{\pi}{\Omega}\left(\frac{1}{2} - \frac{1}{2^{n+1}}\right), &\text{otherwise}
    \end{cases}
    .
\end{align}
We use this specified lattice site $x_{nr_n}$ when studying the effect of varying the scale parameters of the interacting coupling strength as the positions $x_{nj_n}$ are specified so that for any $n$, these are the closest to $1/2$ for a given scale and thus closest to each other as desired when isolating and studying effects dependent only on wavelet scale.

For lattice coupling strengths at the same scale (with all $n_i$ equal), we can evaluate the integrals in Eq.~\ref{quarticoperatorfourierrep} directly to get the exact discrete lattice site couplings for some special cases of choices of the translations $j_i$. 

For the case of three equal $j$ indices and one difference, we have
\begin{align}
    \mathcal{E}^{nnnn}_{j,j,j,j+d} = \frac{1}{h_n}
    \begin{cases}
        \frac{1}{2}, &d=0,\\
        -\frac{3\left(1+(-1)^d\right)}{4\pi^2d^2}, &d\neq 0,
    \end{cases}
    \label{couplingtensor3-1slice}
\end{align}
which leaves $\mathcal{E}_{j,j,j,j+d}=0$ for odd separations $d$ and strictly negative contributions otherwise.
For paired translations $(j,j,j+d,j+d)$ the coupling tensor is given by
\begin{align}
    \mathcal{E}^{nnnn}_{j,j,j+d,j+d} = \frac{1}{h_n}
    \begin{cases}
        \frac{1}{2}, &d=0,
        \\
        \frac{2+(-1)^d}{\pi^2d^2}, &d\neq0.
    \end{cases}
    \label{couplingtensor2-2slice}
\end{align}
Here, we can see that the tensor along the paired separation slice at same scale is strictly positive with inverse quadratic decay in separation, d. 

\begin{figure}[t]
    \centering
    \includegraphics[width=\linewidth]{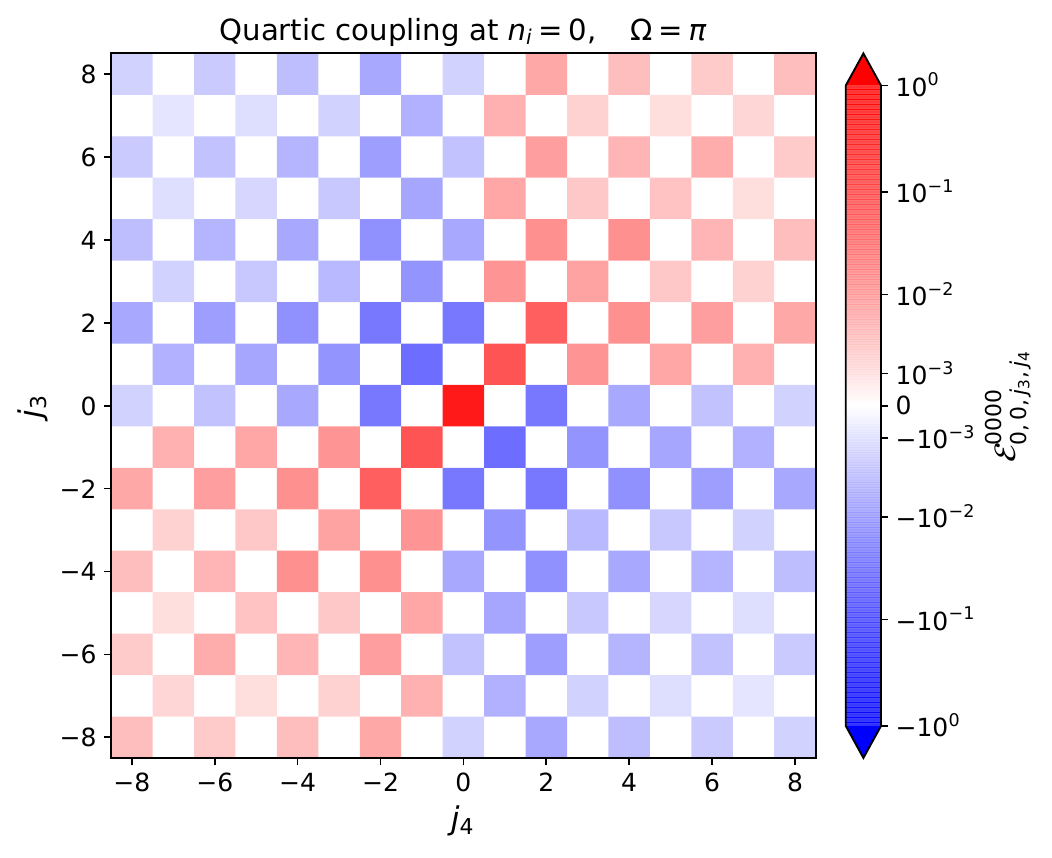}
    \caption{Plot of the quartic coupling strength $\mathcal E^{0000}_{00j_3j_4}$ on lattice sites at the same scale $n=0$ and fixed $j_1=j_2=0$, with $\Omega=\pi$. The orthonormal-wavelet convention is the same as in Fig.~\ref{fig:interactioncouplingplotnsep}. As expected, the coupling strength is strongest at the same lattice site and decays in magnitude for increasing separation.}
    \label{fig:interactioncouplingplotjsep}
\end{figure}

A plot of the same scale coupling tensor for varying $j_3$ and $j_4$ is shown in Fig.~\ref{fig:interactioncouplingplotjsep}. Note that the horizontal and vertical lines centred at the origin represent Eq.~\eqref{couplingtensor3-1slice} and the centre diagonals represent Eq.~\eqref{couplingtensor2-2slice}. The rest of the lattice has been filled in through numerical evaluation of Eq.~\eqref{quarticoperatorfourierrep}. 

\subsection{Lattice coupling strength from interactions on a dyadic rational lattice}
In the dyadic rational co-ordinate basis the quartic interaction Hamiltonian is
\begin{align}
    \op H_I =& \frac{\lambda}{4!}\int_\reals dx \left(\sum_{\eta\in\mathbb D^\times}\op{Q}_\eta u(x;\eta)\right)^4
    \nonumber
    \\
    =&
    \frac{\lambda}{4!}\sum_{\bvec{\eta}\in(\mathbb D^\times)^4}\op{Q}_{\eta_1}\op{Q}_{\eta_2}\op{Q}_{\eta_3}\op{Q}_{\eta_4}\mathcal{E}_{\bvec{\eta}},
\end{align}
where again $\mathcal{E}$ is given by the overlap of the four Wavelets, given by
\begin{align}
    \mathcal{E}_{\bvec{\eta}} \coloneqq \int_\reals dx u(x;\eta_1)u(x;\eta_2)u(x;\eta_3)u(x;\eta_4),
\end{align}
and $\bvec{\eta} \coloneqq (\eta_1, \eta_2,\eta_3,\eta_4)$ simply specifies the four dyadic rational parameters.

It is here that the advantage of considering a dyadic rational basis for wavelet based quantum field theory becomes apparent. In the usual position + scale representation, the coupling strength between field operators is parameterised by eight integers and there is little intuition as to how the coupling coefficients between oscillators on the lattice sites should behave. However, on the lattice of dyadic rationals, this intuition becomes clear while dealing with half of the parameters. We observe that the coupling strength between lattice sites scales with both position distance considering the usual distance norm as well as dyadic distance considering the dyadic rational distance norm.

The two tensors now differ only in their index labels:
\begin{align}
    \mathcal E_{\bvec\eta}
       &=\mathcal E^{n_1n_2n_3n_4}_{j_1j_2j_3j_4}.
    \label{quarticnormalizationdictionary}
\end{align}
For comparison with the cardinal sampling kernels of Sec.~\ref{sec:waveletdecomposition}, define
\begin{align}
    \mathcal E^{(w)\,\bvec n}_{\bvec j}
       &\coloneqq\int_\reals dx\,\prod_{r=1}^4 w^{n_r}_{j_r}(x).
\end{align}
The normalization dictionary then gives
\begin{align}
    \mathcal E^{\bvec n}_{\bvec j}
       &=\frac{\mathcal E^{(w)\,\bvec n}_{\bvec j}}
               {\sqrt{h_{n_1}h_{n_2}h_{n_3}h_{n_4}}}.
\end{align}
No such prefactor is needed when converting between the two orthonormal coordinate representations. In particular, at one scale the cardinal tensor scales as $h_n$ and the orthonormal tensor as $1/h_n$.

Consequently, we have
\begin{align}
    \mathcal{E}_{\bvec{\eta}} =&
    \frac{(2\pi)^2}{\Omega^2\sqrt{|\eta_1|_2|\eta_2|_2|\eta_3|_2|\eta_4|_2}}\int_\reals \frac{dk_1dk_2dk_3}{(2\pi)^3}
    \nonumber
    \\
    &\times e^{-i\Psi(\bvec{k})}\prod_{r=1}^3\chi_{\eta_r}(k_r)\chi_{\eta_4}(-k_1-k_2-k_3),
\end{align}
where $\chi_{\eta_r}$ is given by Eq.~\eqref{dyadicrationalindicatormomentum} and $\Psi(\bvec{k})$ is given by
\begin{align}
    \Psi(\bvec{k}) \coloneqq \frac{\pi}{\Omega}\left(k_1(\eta_1-\eta_4)+k_2(\eta_2-\eta_4)+k_3(\eta_3-\eta_4)\right).
\end{align}
The same support condition can be written on the dyadic rational lattice. For $\mathcal E_{\bvec\eta}$ to be non-zero, there must be a non-empty proper subset $S$ of the four labels such that
\begin{align}
    \frac{1}{2}<\frac{\sum_{i\in S}|\eta_i|_2}{\sum_{j\in S^c}|\eta_j|_2}<2.
    \label{dyadicrationaladmissabilitycondition}
\end{align}
If no such partition exists, the coefficient vanishes for all positions with those dyadic magnitudes. If one does exist, the coefficient is allowed by momentum support but may still vanish because of the phases associated with the specified positions. Thus this is the same necessary support condition as Eq.~\eqref{interactionadmissability}, expressed in dyadic co-ordinates; it does not guarantee a non-zero value for each admissible set of sites.

\begin{figure}[t]
    \centering
    \includegraphics[width=\linewidth]{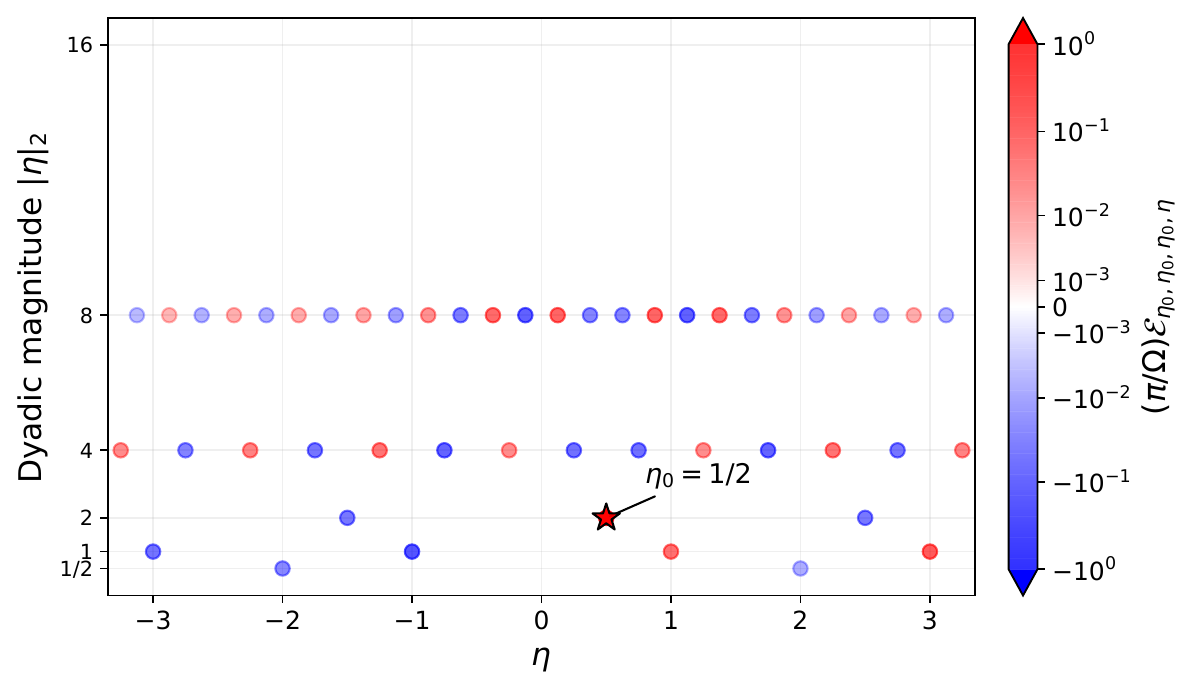}
    \caption{Plot of the dimensionless quartic coupling $(\pi/\Omega)\mathcal E_{\eta_0,\eta_0,\eta_0,\eta}$ on the lattice of dyadic rational numbers. This is a cross-section studying the coupling between the fixed site $\eta_0=1/2$ and the variable site $\eta\in\mathbb D^\times$, with three oscillators at $\eta_0$ and one at $\eta$. The sites have been separated vertically by their dyadic magnitudes, with $\nu_2(\eta)\geq-4$. Note that when $|\eta|_2$ is sufficiently large compared with $|\eta_0|_2=2$, the support condition forces the 3-1 coupling to vanish. Here the row $|\eta|_2=16$ is zero.}
    \label{fig:3-1couplingstrengthdyadictensor}
\end{figure}

\begin{figure}[t]
    \centering
    \includegraphics[width=\linewidth]{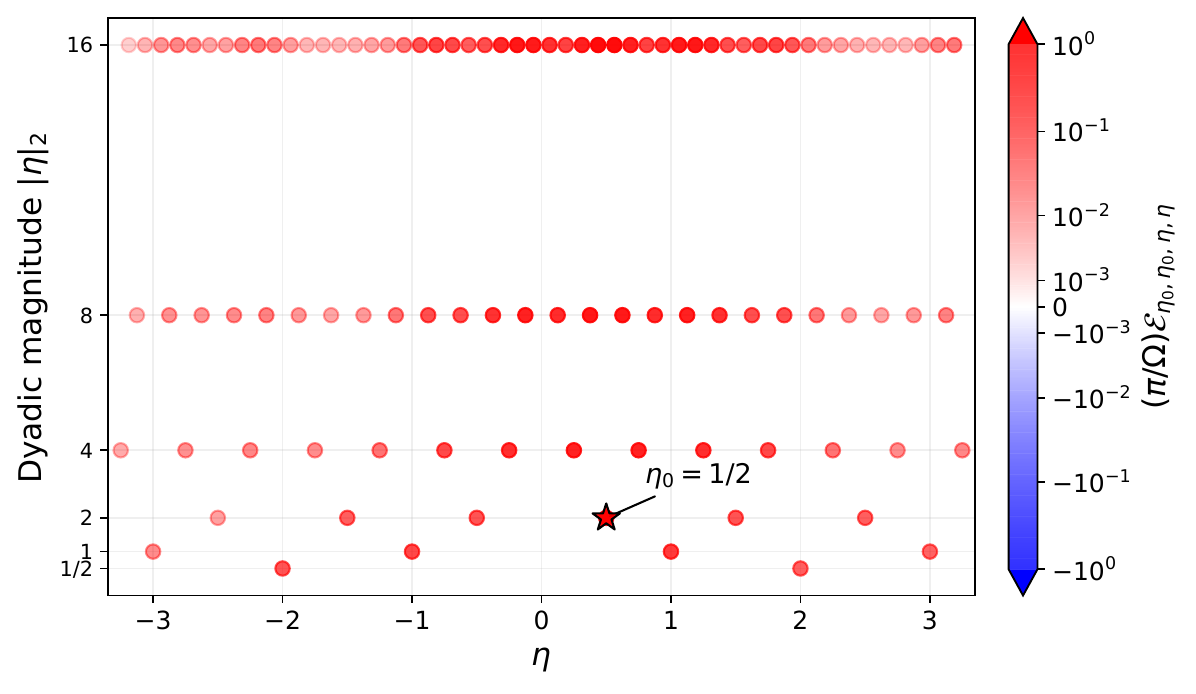}
    \caption{Plot of the dimensionless quartic coupling $(\pi/\Omega)\mathcal E_{\eta_0,\eta_0,\eta,\eta}$ on the lattice of dyadic rational numbers, with the same sites and vertical ordering as Fig.~\ref{fig:3-1couplingstrengthdyadictensor}. This cross-section has pairs of oscillators at $\eta_0=1/2$ and $\eta\in\mathbb D^\times$, respectively. Note that unlike the 3-1 tensor, this paired 2-2 tensor remains strictly positive at arbitrary separation in dyadic magnitude. The origin is not a wavelet site and is omitted.}
    \label{fig:2-2couplingstrengthdyadictensor}
\end{figure}

If we choose $\eta, \eta'$ to have the same dyadic magnitude, i.e., $|\eta|_2 = |\eta'|_2$, then we can study the decay of $\mathcal{E}_{\bvec{\eta}}$ along slices of equal dyadic magnitude, similar to studying the value of the quartic coupling tensor on lattices of the same scale. For $|\eta|_2=|\eta'|_2$, define the rescaled separation
\begin{equation}
    d\coloneqq\frac{|\eta|_2}{2}(\eta-\eta')=\frac{x_\eta-x_{\eta'}}{h_\eta}.
    \label{rescaleddyadicseparation}
\end{equation}
This measures the separation in units of the sampling spacing within that scale and gives a convenient variable for the following coupling coefficients. We have
\begin{subequations}
    \begin{align}
        \mathcal{E}_{\eta,\eta,\eta,\eta'} &= \frac{|\eta|_2\Omega}{2\pi}
        \begin{cases}
            \frac{1}{2},&\eta=\eta',
            \\
            -\frac{3\left(1+(-1)^d\right)}{4\pi^2d^2}&\eta'\neq\eta,
        \end{cases}
        \\
        \mathcal{E}_{\eta,\eta,\eta',\eta'} &=\frac{|\eta|_2\Omega}{2\pi}
        \begin{cases}
            \frac{1}{2}, &\eta=\eta',
            \\
            \frac{2+(-1)^{d}}{\pi^2d^2}, &\eta\neq\eta',
        \end{cases}
        \label{31and22dyadiccoupling}
    \end{align}
\end{subequations}
The same lattice site self coupling strength is therefore given by
\begin{align}
    \mathcal{E}_{\eta\eta\eta\eta} = \frac{|\eta|_2\Omega}{4\pi}. 
\end{align}
We can see then that for increasing dyadic magnitude of a given lattice site, the self coupling strength for the quartic interacting field increases linearly. 
The forms of these slices through the $\mathcal{E}_{\bvec{\eta}}$ tensor match up with their counterparts on the regular $n,j$ Shannon wavelet lattice, as expected.
However, it is important to note that $\mathcal E$ need not be zero for $|\eta|_2\neq|\eta'|_2$. As such, it is useful to study the slices in Eq.~\eqref{31and22dyadiccoupling} for unequal dyadic magnitudes, as in Figs.~\ref{fig:3-1couplingstrengthdyadictensor} and \ref{fig:2-2couplingstrengthdyadictensor}. With $|\eta_0|_2=2$, the choice $|\eta|_2=16$ in the 3-1 slice fails the support condition and therefore gives zero coupling. The origin $\eta=0$ is excluded from the wavelet lattice $\mathbb D^\times$ and does not represent an additional scale.

For the paired 2-2 slice there is a stronger conclusion than support admissibility alone. In this case,
\begin{align}
    \mathcal E_{\eta,\eta,\eta',\eta'}
       &=\int_\reals dx\,u(x;\eta)^2u(x;\eta')^2>0.
    \label{pairedquarticpositivity}
\end{align}
The integrand is non-negative, and two non-zero Shannon wavelets are simultaneously non-zero on an interval. Consequently, this particular paired coefficient is strictly positive for any two wavelet sites, even at widely separated scales. Its non-vanishing follows from the position-space integral, whereas the general selection rule remains only a necessary support condition.

\begin{figure}[t]
    \centering
    \includegraphics[width=\linewidth]{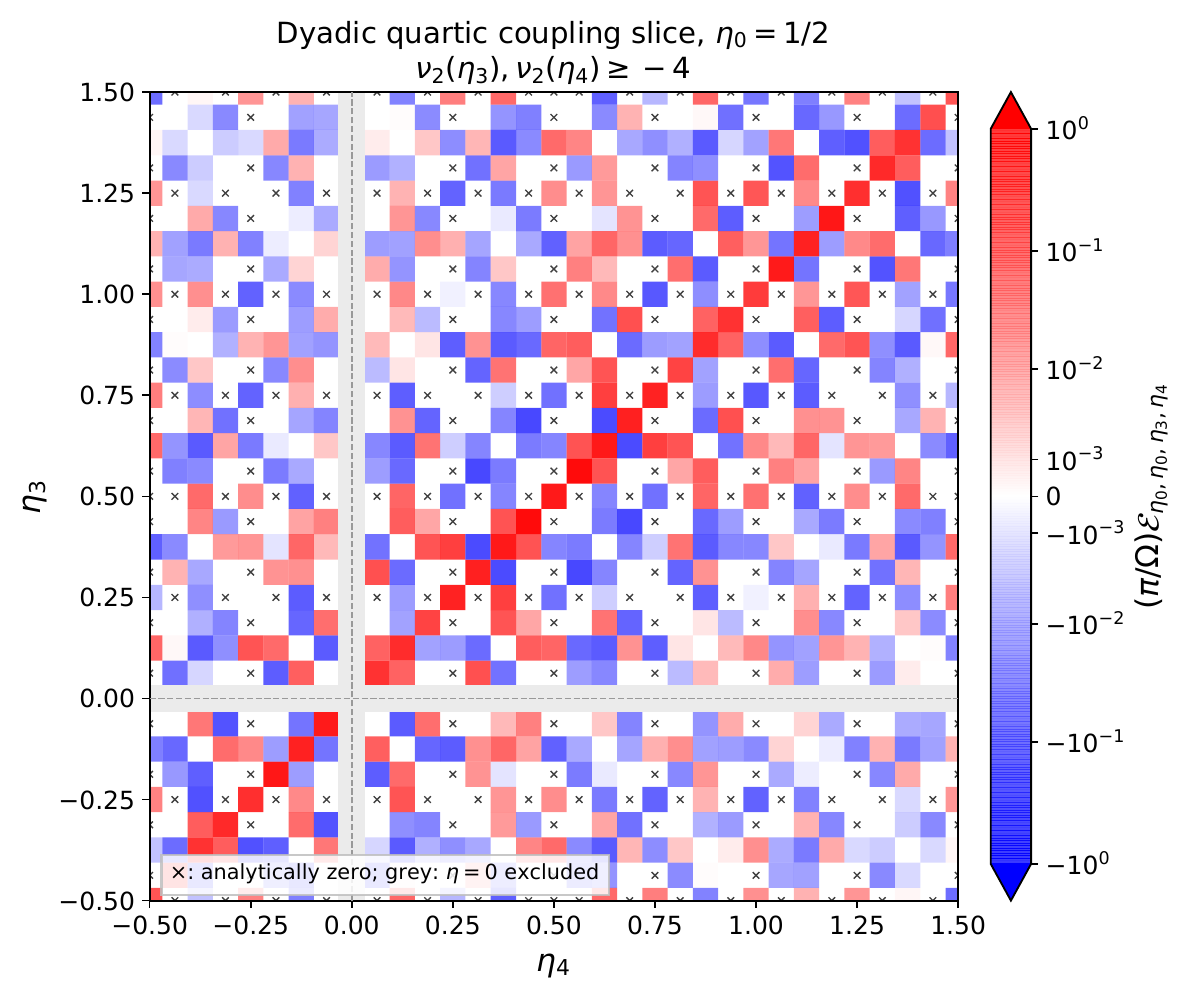}
    \caption{Plot of the dimensionless quartic interaction tensor $(\pi/\Omega)\mathcal E_{1/2,1/2,\eta_3,\eta_4}$ for varying $\eta_3$ and $\eta_4$ between $-0.5$ and $1.5$. The density of the lattice has been truncated at $|\eta_i|_2\leq16$, or equivalently $\nu_2(\eta_i)\geq-4$, for clarity. The plotted interval is centred at $(1/2,1/2)$ to show the symmetric structure of this tensor. Crosses mark analytically established zeros, while grey bands with dashed centre lines indicate the excluded site $\eta=0$.}
    \label{fig:3-1-1couplingstrengthdyadictensorlattice}
\end{figure}
\begin{figure}[t]
    \centering
    \includegraphics[width=\linewidth]{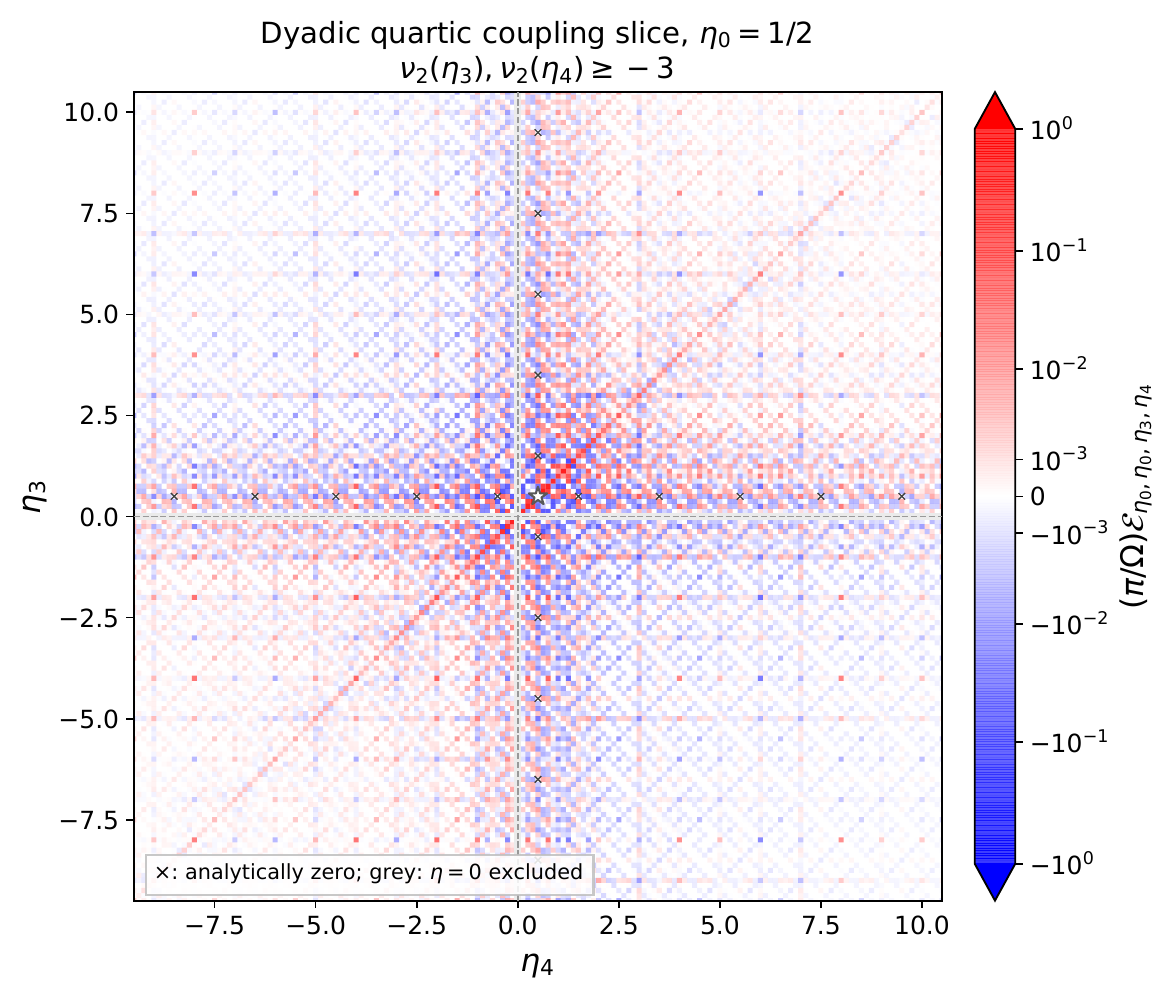}
    \caption{Plot of the dimensionless quartic interaction tensor $(\pi/\Omega)\mathcal E_{1/2,1/2,\eta_3,\eta_4}$ for varying $\eta_3$ and $\eta_4$. The density of the lattice has been truncated at $|\eta_i|_2\leq8$, or equivalently $\nu_2(\eta_i)\geq-3$, for clarity. The range along both axes has been extended to $[-9.5,10.5]$ to show the decay in coupling strength as real distance grows. Crosses and grey bands have the same meaning as in Fig.~\ref{fig:3-1-1couplingstrengthdyadictensorlattice}, and the star marks $(1/2,1/2)$.}
    \label{fig:3-1-1couplingstrengthdyadictensorlatticelongrange}
\end{figure}
\begin{figure*}[t]
    \centering
    \includegraphics[width=\linewidth]{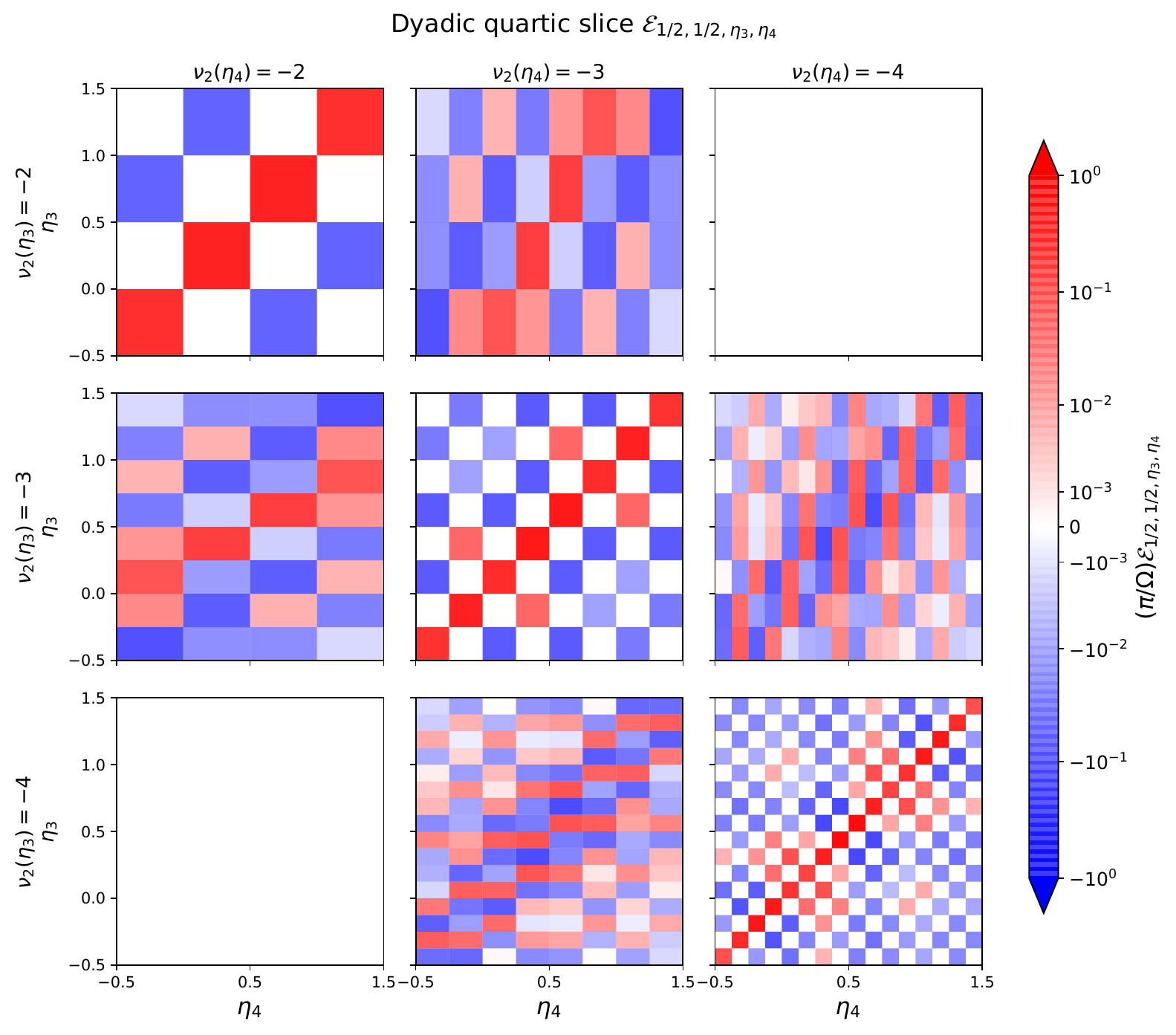}
    \caption{Plots of the dimensionless quartic coupling tensor $(\pi/\Omega)\mathcal E_{1/2,1/2,\eta_3,\eta_4}$ on the dyadic rational lattice, with the sites grouped by their exact valuations. These panels show a subset of the coupling data in Fig.~\ref{fig:3-1-1couplingstrengthdyadictensorlattice}, regrouped by valuation. Rows specify $\nu_2(\eta_3)$ and columns specify $\nu_2(\eta_4)$, each taking the values $-2,-3,-4$. The diagonal panels therefore have equal dyadic magnitudes $4$, $8$ and $16$, from top left to bottom right. The off-diagonal panels show mixed magnitudes; for instance, the top centre panel has $|\eta_3|_2=4$ and $|\eta_4|_2=8$. Reading the diagonal from bottom right to top left decreases the dyadic magnitude and gives a coarser sampling lattice. The bottom left and top right panels vanish because their scale configurations fail the support condition in Eq.~\eqref{dyadicrationaladmissabilitycondition}.}
    \label{fig:quartictensorsepr}
\end{figure*}

We can observe further the behaviour of $\mathcal{E}_{\bvec{\eta}}$ on the dyadic rational lattice by plotting its values for $\mathcal{E}_{1/2,1/2,\eta_3,\eta_4}$ to observe how differing lattice sites are coupled together. This can be seen in Fig.~\ref{fig:3-1-1couplingstrengthdyadictensorlattice}. The density of the lattice here is truncated at $\nu_2(\eta) \geq -4$ or $|\eta_i|_2 \leq16$ to give a finite and interpretable lattice. We can see here that difference in dyadic magnitude between indices plays a large role in determining the coupling strength between dyadic rational lattice sites through Eq.~\eqref{dyadicrationaladmissabilitycondition}. We also see on average the strength of the coupling between lattice sites decrease with real distance between the lattice sites as well. This can be further observed in Fig.~\ref{fig:3-1-1couplingstrengthdyadictensorlatticelongrange} which shows a similar plot but over a far greater real distance for $\eta_3$ and $\eta_4$ but with a slightly lower density of lattice sites. Here, the quartic coupling tensor can be clearly seen to approach zero as the real distance between lattice sites increases.
Given that, we can see that both real distance between lattice sites $|\eta-\eta'|$ as well as difference in dyadic magnitude $|\eta|_2/|\eta'|_2$ govern the coupling strength of quartic interactions on a lattice of dyadic rationals. 
It is also useful to observe how the quartic coupling tensor behaves as one proverbially zooms in or out in scale or dyadic valuation. This is shown in Fig.~\ref{fig:quartictensorsepr}. We can see that the lattice sites/scales of the wavelet basis act as a form of coarse graining of $\mathcal{E}_{\bvec{\eta}}$ along the block diagonal of the plots in Fig.~\ref{fig:quartictensorsepr} while the off diagonal show the mixed scale couplings of the quartic tensor. This is finally exacerbated in the top right and bottom left plots which show the full decoupling of the lattice sites through $\mathcal{E}$ as the isolated scale difference becomes too large and the admissibility condition in Eq.~\eqref{dyadicrationaladmissabilitycondition} is no longer satisfied.

\section{Correlation functions for interacting fields in a Shannon wavelet basis \label{sec:interactingcorrelationfunctionswavelets}}
When studying quantum field theory from an information theoretic approach one of the more commonly used tools is the vacuum two-point correlation function. For a non-interacting free scalar field this is of course, given by
\begin{align}
    K(x,t;x',t')\coloneqq\bra{0}\op{\phi}(x', t')\op{\phi}(x, t)\ket{0},
\end{align}
which measures vacuum correlations between the fields at the two spacetime points. As discussed in the previous sections, the free-field coefficients $\op Q^n_j(t)$ and $\op Q_\eta(t)$ have non-zero two-point correlations only within one scale, or equivalently between sites with the same dyadic magnitude. It is therefore pertinent to ask how this property is modified by interactions.

For perturbative calculations we use the time-ordered Feynman Green's function
\begin{align}
    G_F(x,t;x',t')&\coloneqq
       \bra{\Omega}\mathcal T\{\op\phi(x,t)\op\phi(x',t')\}\ket{\Omega},
\end{align}
where $\ket{\Omega}$ is the interacting vacuum. This differs from the unordered function $K$ above; our definition of $G_F$ includes no additional overall factor of $i$. In the interaction picture, define the quartic Hamiltonian density and its spatial integral by
\begin{align}
    \op{\mathcal H}_I(z,\tau)&\coloneqq
       \frac{\lambda}{4!}\op\phi_I(z,\tau)^4,\\
    \op H_I(\tau)&\coloneqq\int_\reals dz\,\op{\mathcal H}_I(z,\tau).
    \label{interactionpicturedensity}
\end{align}
The Feynman function can then be written in terms of free-vacuum expectation values as
\begin{align}
    &G_F(x,t;x',t')
    \nonumber\\
       &\quad=\frac{\bra0\mathcal T\left\{
          \op\phi_I(x,t)\op\phi_I(x',t')
          e^{-i\int_\reals d\tau\,\op H_I(\tau)}
          \right\}\ket0}
          {\bra0\mathcal T\left\{
          e^{-i\int_\reals d\tau\,\op H_I(\tau)}
          \right\}\ket0}.
    \label{interactingvacuumexpansion}
\end{align}
Here the time ordering acts on the external fields and all interaction vertices together, and the denominator removes disconnected vacuum bubbles. We assume the usual adiabatic vacuum prescription: the interaction is switched on and off with a factor $e^{-\epsilon_{\mathrm{ad}}|\tau|}$, with $\epsilon_{\mathrm{ad}}>0$ taken to zero after forming the ratio, and the free propagators carry the Feynman $i0$ prescription. The interacting vacuum is taken to be continuously connected to the free vacuum in this perturbative expansion.

The fields themselves in the interaction picture are given by
\begin{align}
    \op{\phi}_I(x, t) \coloneqq \int_\reals \frac{dk}{2\pi}\frac{1}{\sqrt{2\omega_k}}\op{a}_ke^{ikx - i\omega_k t} + \text{H.c.}
\end{align}
As such, $\op{\phi}_I$ in the interaction picture is simply the free field after it has evolved in time under the free Hamiltonian.

\subsection{Two-point correlation functions between Shannon wavelet transformed fields}
For fields under the Shannon wavelet decomposition, the unitary that governs time evolution for the interacting fields in this picture is given by
\begin{subequations}
\begin{align}
    \op{U}(t,t') &\coloneqq\mathcal T e^{-i\int_{t'}^t d\tau\,\op H_I(\tau)}
    \nonumber
    \\*
    &=
    \mathcal{T}e^{-i\frac{\lambda}{4!}{\int_{t'}^t}d\tau \sum_{(\bvec n,\bvec j)\in\mathbb Z^8}\prod_{k = 1}^4\left[\op{Q}^{n_k}_{I,j_k}(\tau)\right]\mathcal{E}_{\bvec{j}}^{\bvec{n}}},
    \\*
    &=
    \mathcal{T}e^{-i\frac{\lambda}{4!}{\int_{t'}^t}d\tau \sum_{\bvec{\eta}\in(\mathbb D^\times)^4}\prod_{k=1}^4\left[\op{Q}_{I,\eta_k}(\tau)\right]\mathcal{E}_{\bvec{\eta}}},
\end{align}
\label{interactingevolutionop}
\end{subequations}
in the $(n,j)$ and dyadic lattice representations, respectively. Here $\op Q^n_{I,j}(t)\coloneqq\int dx\,\overline{u^n_j(x)}\op\phi_I(x,t)$ and $\op Q_{I,\eta}=\op Q^n_{I,j}$; the interaction carries the same $\lambda/4!$ in both representations.
It is useful to note that while it can be written in terms of Shannon wavelet transformed fields, this unitary is itself unchanged by rerepresentation in either Shannon wavelet basis so long as there are no truncations in $(\bvec{n},\bvec{j})$ or $(\bvec{\eta}, \nu_2(\bvec{\eta}))$, respectively. If any of these parameters are cut-off then the respective sums over their indices will be cut-off and Eqs.~\eqref{interactingevolutionop} will be reduced to an approximation of the continuous interacting time evolution operator.

The Feynman function in the two lattice representations is therefore
\begin{subequations}
\begin{align}
    &G_F(n,j,t;n',k,t')
    \nonumber\\
       &\quad=\frac{\bra0\mathcal T\left\{
           \op Q^n_{I,j}(t)\op Q^{n'}_{I,k}(t')
           \op U(\infty,-\infty)\right\}\ket0}
           {\bra0\op U(\infty,-\infty)\ket0},\\
    &G_F(\eta,t;\eta',t')
    \nonumber\\
       &\quad=\frac{\bra0\mathcal T\left\{
           \op Q_{I,\eta}(t)\op Q_{I,\eta'}(t')
           \op U(\infty,-\infty)\right\}\ket0}
           {\bra0\op U(\infty,-\infty)\ket0}.
\end{align}
\end{subequations}
As in Eq.~\eqref{interactingvacuumexpansion}, $\op U(\infty,-\infty)$ is to be expanded under the same time-ordering operation as the external fields, with the same adiabatic prescription. These lattice functions correlate the canonical coefficients $Q$, whereas $G_F(x,t;x',t')$ with continuous spatial arguments correlates $\phi$. A projected-field sample correlator is obtained by dividing the corresponding $Q$ correlator by $\sqrt{h_nh_m}$, or $\sqrt{h_\eta h_{\eta'}}$ in dyadic co-ordinates.

Now, it is important to note that the two-point function is translationally invariant on spacetime and, as such, observes momentum conservation. We can see this through its mixed spatial momentum/time representation as
\begin{align}
    \tilde{G}_F(k,k';t,t') &= \bra{\Omega}\mathcal{T}\op{\tilde{\phi}}(k;t)\op{\tilde{\phi}}(k';t')\ket{\Omega}
    \nonumber
    \\
    &= 2\pi\delta (k+k') \tilde{G}_F(k;t,t').
\end{align}
For fields separated in the Shannon wavelet basis, then, the two-point correlation function between lattice sites can be written as
\begin{align}
    &\bra{\Omega}\mathcal{T}\op{Q}^n_j(t)\op{Q}^m_h(t')\ket{\Omega}
       \nonumber\\
       &\quad=\int_\reals\frac{dk}{2\pi}
          \overline{\widetilde u^n_j(k)}\,
          \overline{\widetilde u^m_h(-k)}\,\tilde G_F(k;t,t')
       \nonumber\\
       &\quad=\sqrt{h_nh_m}\int_\reals\frac{dk}{2\pi}
          e^{ik(x_{nj}-x_{mh})}
          \chi_n(k)\chi_m(k)\tilde G_F(k;t,t').
    \label{waveletreptwopointinteractingfunction}
\end{align}
For distinct scales $n\neq m$ these wavelets have no overlapping support and as a result, the two-point function between scales is exactly zero. Subsequently, the same must be true for two-point correlation functions between lattice sites of differing dyadic magnitude in the dyadic rational representation of the Shannon wavelet lattice. Explicitly, we have for $n\neq m$ or, equivalently, $|\eta|_2\neq |\eta'|_2$,
\begin{subequations}
    \begin{align}
        G_F(n,j,t; m,h,t') &= 0,
        \\
        G_F(\eta,t;\eta',t') &=0.
    \end{align}
    \label{interactingtwopointfunctionsarezero}
\end{subequations}
This is true non-perturbatively for the interacting vacuum $\ket{\Omega}$. So, despite the fact that the quartic term in the Hamiltonian for the field acts to couple the different scales in the Shannon wavelet lattice together, the two-point function is blind to this coupling regardless of the presence of interactions. This occurs because despite the ability for internal self-energy diagrams such as a tadpole diagram to redistribute momentum among their internal lines, the net momentum for the external lines must sum to zero and maintain momentum conservation. Since the Shannon wavelet scales share no momentum overlap in their spectra, there can be no non-zero two point functions between the scales, despite the interactions coupling these scales together. 

That being said, we can observe how corrections to the two-point correlator from interactions behave if we restrict to only same scale parameterisations of the co-ordinates. In that we require $n=m$ or, equivalently on the dyadic rational lattice $|\eta|_2=|\eta'|_2$. 

At equal time and scale, write $G_F(n,j-h;t=t')$ for the canonical mode correlator at scale $n$. In dyadic coordinates we retain both site labels, since their common magnitude as well as their separation determines the normalization. The two-point function reduces to
\begin{subequations}
\begin{align}
    G_F(n,j-&h; t=t')
    \nonumber
    \\
    &= \frac{1}{2^n\Omega}\int_{2^n\Omega}^{2^{n+1}\Omega}dk \cos\left(2^{-n}\frac{k\pi}{\Omega}(j-h)\right) C(k)
    \\
    G_F(\eta,&\eta';t=t')
    \nonumber
    \\&=\frac{2}{|\eta|_2\Omega}\int_{|\eta|_2\Omega/2}^{|\eta|_2\Omega}dk\cos\left(k\frac{\pi}{\Omega}(\eta-\eta')\right)C(k),
\end{align}
\label{equaltimetwopointinteractingcorrelators}
\end{subequations}
where $C(k)$ is the equal time kernel for the propagator and is given by
\begin{align}
    C(k) \coloneqq \tilde{G}_F(k;t=t') = \int_\reals \frac{dk_0}{2\pi}\mathcal{G}_F(k_0, k),
\end{align}
with $\mathcal{G}_F$ being the fully Fourier transformed Feynman propagator.
Corrections to $G_F$ due to interactions can then be determined via perterbative corrections to $C(k)$. 
Under this convention for the Feynman function, the free momentum-space propagator and its equal-time kernel are
\begin{align}
    \mathcal G_F^{(0)}(k_0,k)
       &=\frac{i}{k_0^2-k^2-m^2+i0},\\
    C_0(k)&\coloneqq\int_\reals\frac{dk_0}{2\pi}\mathcal G_F^{(0)}(k_0,k)
       =\frac{1}{2\omega_k}.
    \label{freefeynmanequaltimekernel}
\end{align}
Here $\omega_k=\sqrt{k^2+m^2}$, as it is in the free theory. The same-scale free-field correlators are consequently
\begin{subequations}
\begin{multline}
    G_F^{(0)}(n,j-h;t=t')\\
       =\frac{1}{2a_n}\int_{a_n}^{2a_n}dk\,
          \frac{\cos\bigl(kh_n(j-h)\bigr)}{\sqrt{k^2+m^2}},
\end{multline}
\begin{multline}
    G_F^{(0)}(\eta,\eta';t=t')\\
       =\frac{1}{|\eta|_2\Omega}
         \int_{|\eta|_2\Omega/2}^{|\eta|_2\Omega}dk\,
         \frac{\cos\bigl(k(x_\eta-x_{\eta'})\bigr)}{\sqrt{k^2+m^2}},
\end{multline}
\end{subequations}
where the second expression assumes $|\eta|_2=|\eta'|_2$.
From here, one can calculate the corrections to the free field correlation function due to interactions via a perterbative expansion of the Dyson series~\cite{UpdatedPeskin2018}. For instance, to first order in $\lambda$, we can write the intial `one-loop' correction to $G_F^{(0)}$ as
\begin{align}
    &\delta G_F^{(1)}(x,t;y,t')
       \nonumber\\
       &\quad=-\frac{i\lambda}{2}G_{F}^{(0)}(0,0)
          \int_\reals d\tau\int_\reals dz\,
       \nonumber\\
       &\qquad\quad\times
          G_F^{(0)}(x-z,t-\tau)G_F^{(0)}(z-y,\tau-t').
    \label{tadpolepositioncorrection}
\end{align}
After a choice of regulator for the loop component of Eq.~\eqref{tadpolepositioncorrection} given by $G_F^{(0
0)}$ one could then calculate explicitly the corrections to the two-point function due to quartic interactions up to order $\lambda$. However, given that such a regulating cut-off need not depend on any given wavelet scale due to the scale-independence of the interacting time-evolution unitary, we leave this explicit calculation and calculation of the corrections at higher orders of $\lambda$ to future work. Suffice it to say that given the free-field \emph{same-scale} two-point functions are not ubiquitously zero and the loop component is also non-zero, we can see that the first order correction to the two-point function is also non-zero.

Given this, we can observe that despite the quartic interactions that act to couple the wavelet scales together, the vacuum two-point function remains `scale-diagognal'. Meaning that the only non-zero components of the two-point function in the Shannon wavelet decomposition will be those with the start and end-points on equal scales, regardless of interactions.

\subsection{\texorpdfstring{$N$}{N}-point functions for cross scale interactions}
Given that the cross-scale coupling of the lattice sites from interactions is invisible to the vacuum two-point function, it is useful to determine what objects can observe this coupling. For this means we turn our attention to higher order correlations. Namely, the $N$-point function. Such an object is useful for studying processes such as scattering. We can write this $N$-point function as%
\footnote{Note that in this section we will not be dealing with perturbative expansions of the correlation function. As such, the superscript notation $^{(N)}$ denotes only the fact that there are $N$ external fields present in the correlation function, with nothing to do with the notational labels for the terms in the perturbative expansion on the two-point Feynman propagator in the previous section.}%
\begin{align}
    G^{(N)}(x_1, t_1, ...,x_N, t_N) &\coloneqq \bra{\Omega}\mathcal{T}\prod_{p=1}^N\op{\phi}(x_p, t_p)\ket{\Omega}.
\end{align}
Given that the $N$-point function remains translation invariant and thus maintains total momentum conservation across the external fields, we can make note that
\begin{align}
    \tilde{G}^{(N)}(k_1, ...,&k_N, t_1, ..., t_N) 
    \nonumber
    \\*
    &= \bra{\Omega}\mathcal{T}\prod_{p=1}^N \op{\tilde{\phi}}(k_p, t_p)\ket{\Omega}
    \nonumber
    \\*
    &= 
    2\pi \delta\left(\sum_{p=1}^Nk_p \right)\tilde{G}(k_1, ..., k_N, t_1, ..., t_N).
\end{align}
where $\tilde{G}^{(N)}$ is the partial Fourier transform of the $N$-point function with respect to only the position labels $x_p$.
When we replace the external fields on the real line with their Shannon wavelet representative counterparts on the $(n,j)$ or $\mathbb D^\times$ lattices, we can express this $N$-point function in terms of the wavelets themselves, similar to what was done in Eq.~\eqref{waveletreptwopointinteractingfunction}, as
\begin{align}
    &\bra{\Omega}\mathcal{T}\prod_{p=1}^N\op{Q}^{n_p}_{j_p}(t_p)\ket{\Omega}
    \nonumber\\
    &\quad= \int_{\reals^N}\prod_{p=1}^N
       \left[\frac{dk_p}{2\pi}\overline{\widetilde u^{n_p}_{j_p}(k_p)}\right]
       \,2\pi\delta\left(\sum_{p=1}^N k_p\right)
    \nonumber\\
    &\qquad\quad\times\tilde{G}(k_1, ..., k_N, t_1, ..., t_N).
\end{align}
We can see here that the $N$-point function, due to translation invariance and thus momentum conservation, possesses a delta function in momentum that requires all external momenta $k_p$ to sum to zero in order for the $N$-point function to be non-zero. When the fields are sampled into their respective wavelet basis, then, the necessary condition for non-zero $N$-point correlation becomes
\begin{align}
    \exists k_p:\quad \chi_{n_p}(k_p)=1\ \text{for all }p,
       \qquad \sum_{p=1}^N k_p=0.
\end{align}
In other words, in order for the momentum conserving Dirac delta function in external momenta to be satisfied, the respective wavelet scales for the input external fields must be close enough to each other such that their respective Fourier supports contain contributing modes that sum to zero. For the special case of $N=2$, the only scenario for which this can be satisfied is for when $n_1=n_2$ and the scales for both external fields are the same, giving the condition that the two-point function in the Shannon wavelet basis is only non-zero when the external fields are at the same scale (or at lattice sites of the same dyadic magnitude $|\eta_1|_2 = |\eta_2|_2$ on the $\mathbb D^\times$ lattice).

Now, similar to what was done with the scale admissibility selection rule regarding coupled lattice sites from the quartic interaction tensor in Eq.~\eqref{interactionadmissability}, we can play the same game and note that for a set of modes $k_p$ summed over in the Dirac delta in the $N$-point function, a subset of these modes $S$ will point in one direction (positive $k_p$) and its complement will point in the other direction (negative $k_p$). The total of these modes will simply be the sum of the elements in both of these sets. We have
\begin{align}
    \delta\left(\sum_{p=1}^Nk_p\right) = \delta\left(\sum_{p \in S}k_p + \sum_{p \notin S}k_p\right). 
\end{align}
Now, given that the support of any given wavelet is given by
\begin{align}
    a_{n_p}<|k_p|<2a_{n_p},
\end{align}
the ranges of the sums of positive or negative pointing modes are given by
\begin{subequations}
\begin{align}
    \Omega\sum_{p\in S}2^{n_p} < \sum_{p\in S}|k_p| < 2\Omega\sum_{p\in S}2^{n_p},
    \\
    \Omega\sum_{p\notin S}2^{n_p} < \sum_{p\notin S}|k_p| < 2\Omega\sum_{p\notin S}2^{n_p}.
\end{align}
\end{subequations}
In order for the total sum over $k_p$ to equal zero, these ranges must overlap, giving the selection rule
\begin{align}
    &\exists S:~\textrm{max}\left(\sum_{p\in S}2^{n_p}, \sum_{p\notin S}2^{n_p}\right)
    \nonumber\\
    &\qquad<2\textrm{min}\left(\sum_{p\in S}2^{n_p}, \sum_{p\notin S}2^{n_p}\right).
\end{align}
Equivalently, we have the condition that there exists a choice of mode directions $S$ such that
\begin{align}
    \frac{1}{2}<\frac{\sum_{p\in S}2^{n_p}}{\sum_{q\notin S}2^{n_q}}<2.
    \label{scaleseparationrangecondition}
\end{align}
If this condition cannot be satisfied for a given choice of scales $n_p, n_q$ and $S$, then the $N$-point correlation function for fields at those scales is kinematically disallowed and must be zero, always.
In the dyadic rational representation of the wavelet basis, this condition simply becomes
\begin{align}
    \exists S:~\frac{1}{2}<\frac{\sum_{p\in S}|\eta_p|_2}{\sum_{q\notin S}|\eta_q|_2} < 2.
\end{align}
For external fields on dyadic rational lattice sites $\eta_p$, if the dyadic magnitudes of these lattice sites are too different from one another then correlation between fields at these sites will also be kinematically disallowed, giving $G^{(N)}(\eta_1, ..., \eta_N, t_1, ..., t_N) = 0$. We can see that when $N=4$, these conditions for admissibility of the $N$-point function are exactly equivalent to the necessary support condition for the quartic coupling tensor in Eq.~\eqref{interactionadmissability}.
It is important to note now that this rule is necessary but not generally sufficient to produce non-zero values of the $N$-point correlation function between scales or sites of different dyadic magnitude. An allowed correlator that permits a valid $S$ may still vanish because of additional phase cancellations or symmetries of the field itself. For instance, for a quartic interacting field theory, all odd-$N$ vacuum correlators vanish, independent of any wavelet momentum support rules.

Now, this partition can be used to produce a bound for the correlation range between \emph{isolated} scales. Consider an external field lies at a unique scale $n_*$ while all other fields exist at scales at or below the next highest scale, $n'$ I.e., we have
\begin{align}
    n_* >n' \geq n_i, ~&\forall i\in1, ..., N-2  .
\end{align}
The smallest available magnitude for a given mode at scale $n_*$ is thus given by
\begin{align}
    |k_*|>2^{n_*}\Omega,
\end{align}
and the largest available magnitude for the sum of the rest of the modes at the other external scales is bound by
\begin{align}
    \sum_{i=1}^{N-2}|k_i| + |k'| < (N-1)2^{n'+1}\Omega.
\end{align}
Momentum conservation for an interaction involving these modes is therefore impossible unless
\begin{align}
    2^{n_*}\Omega <(N-1)2^{n'+1}\Omega.
\end{align}
Thus, for scale difference $\Delta n=n_*-n' \geq 1$, we determine the rule
\begin{subequations}
\begin{align}
    2^{\Delta n}&<2(N-1),
    \\
    \Delta n &\leq \lceil\log_2(N-1)\rceil. 
\end{align}
\label{scalerangerule}
\end{subequations}
This gives a useful rule concerning the minimum proximity of the scales of the external fields in the correlator. For $N$ external fields, a unique highest frequency field cannot sit arbitrarily far above all of the the others. At least one other external field must lie within the logarithmic range $\Delta n$ given in Eq.~\eqref{scalerangerule}. In terms of the dyadic rational lattice, these rules become, for a given external dyadic lattice site $\eta_*$ with dyadic magnitude $|\eta_*|_2$ greater than that of the next highers $|\eta'|_2$, we have
\begin{subequations}
    \begin{align}
        \frac{|\eta_*|_2}{|\eta'|_2}&<2(N-1),
        \\
        \nu_2(\eta') - \nu_2(\eta_*) &\leq \lceil \log_2(N-1)\rceil.
    \end{align}
    \label{dyadicscalerangerule}
\end{subequations}
This corresponds to the condition that a lone field on a finer grained sublattice of $\mathbb D^\times$ corresponding with a higher dyadic magnitude can only be correlated with fields at coarser sublattices so long as the difference in respective lattice density is not too large. The dyadic valuation of the field at one lattice site cannot be too far from the dyadic valuation of at least one of the other external fields at the next densest sublattice.

These rules can also be inverted to give the minimum number of external fields $N$ necessary to have non-zero correlation between one field at an isolated scale and the rest of the $N-1$ external fields. We have, for both representations of the Shannon wavelet lattice
\begin{subequations}
\begin{align}
    &N_{\textrm{min}}(\Delta n) = 2^{\Delta n-1}+2,~&\Delta n\geq 1,
    \\
    &N_{\textrm{min}}\left(\frac{|\eta_*|_2}{|\eta'|_2}\right) = \frac{1}{2}\frac{|\eta_*|_2}{|\eta'|_2}+ 2, ~~&\frac{|\eta_*|_2}{|\eta'|_2}\geq 1.
\end{align}
\label{minimumexternalfieldcondition}
\end{subequations}
In this case, if one wanted to have non-zero correlation between a field at one scale with a difference of $\Delta n$ between that and another field at the next highest scale, the minimum number of external fields $N$ required for $G^{(N)}$ to be non-zero is given by Eqs.~\eqref{minimumexternalfieldcondition}%
\footnote{Note here that this is the minimum number of external fields necessary for kinematically allowed correlations between scales as set by scale differences in the wavelet decomposition. It does not include additional restrictions to correlation functions set by the theory itself. For instance, N=3 is also disallowed by the theory due to Wick's theorem and so for a scale difference of 1 between isolated scales, the minimum number of external fields set by the scale selection rule is 3 but due to the nature of the quartic interactions themselves, the true minimum number of external fields is actually 4.}.%
It is important to note that that Eqs.~\eqref{scalerangerule} and~\eqref{dyadicscalerangerule} are bounds on a \emph{unique isolated highest scale}. There is not necessarily a universal maximum value on $n_{\textrm{max}}-n_{\textrm{min}}$ or, equivalently, $\frac{|\eta|_{2~\textrm{max}}}{|\eta|_{2~\textrm{min}}}$. For example, a four-point correlator with two high-scale legs and two low-scale legs may still be kinematically allowed even for an arbitrarily large scale separation due to the fact that two high frequency momenta pointing in opposite directions may cancel each other while two low frequency momenta pointing in opposite directions do the same, allowing the sum of all momenta to still sum to zero. This is seen most clearly in the fact that the quartic coupling tensor selection rule allows pairs of scales to give non-zero coupling to arbitrarily large ranges in scale distance. The scale-gap rule thus only becomes restrictive when there is one scale sector that must be balanced collectively by a finite number of external fields in the other sectors.

\subsection{Covariance between squared fields as an example of interacting cross scale coupling}
A clear demonstration of the necessity of both interactions as well as higher powers of the field for non-zero cross-scale coupling between wavelet scales is the covariance between quadratic powers of the field at two given scales. Consider
\begin{align}
   C_{AB}\coloneqq \langle A^2 B^2\rangle -\langle \op{A}^2\rangle\langle\op{B}^2\rangle,
\end{align} 
where the operators $\op{A}$ and $\op{B}$ are defined as the time dependent mode variables evaluated at $t=0$ at scale-position co-ordinates $(n,j)$ and $(m,h)$, respectively.
\begin{align}
    \op{A}\coloneqq \op{Q}_j^n(0), 
     ~~~\op{B}\coloneqq \op{Q}_h^m(0).
\end{align}
For a non-interacting field with only quadratic Hamiltonian, we can apply Wick's theorem directly to give
\begin{align}
    C_{AB}^{(0)} 
    &=
    \langle \op{A}^2\op{B}^2\rangle_0 -\langle \op{A}^2\rangle_0\langle\op{B}^2\rangle_0
    \nonumber
    \\
    &=
    2\langle \op{A}\op{B}\rangle_0^2 + \langle \op{A}^2\rangle_0\langle\op{B}^2\rangle_0 - \langle \op{A}^2\rangle_0\langle\op{B}^2\rangle_0
    \nonumber
    \\
    &= 
     2\langle \op{A}\op{B}\rangle_0^2.
\end{align}
This is nothing more than double the square of the free field two-point function and for unequal scales $n\neq m$, this must be equal to zero due to the lack of overlap of Fourier support overlap.
That being said, if we consider this quantity in the interacting vacuum, this quantity gains nonvanishing overlap between wavelet scales at order $\mathcal{O}(\lambda)$. We can see this by evaluating the first contributing terms from the Dyson series. We have
\begin{align}
    C_{AB} 
    &=
     \langle \op{A}^2\op{B}^2\rangle_{\Omega} -\langle \op{A}^2\rangle_{\Omega}\langle\op{B}^2\rangle_{\Omega},
     \label{interactingquadraticfieldcovariance}
\end{align}
The first term of this expression is given by
\begin{align}
    \langle\op{A}^2&\op{B}^2\rangle_\Omega =
    \langle \op{A}^2\op{B}^2\rangle_0 
    \nonumber
    \\
    &- \frac{i\lambda}{4!}\int_\reals d\tau \int_\reals dz \left[\langle\mathcal{T}\op{A}^2\op{B}^2\op{\phi}^4\rangle_0 - \langle\op{A}^2\op{B}^2\rangle_0\langle \mathcal{T}\op{\phi}^4\rangle_0\right] 
    \nonumber
    \\
    &+ \mathcal{O}(\lambda^2),
\end{align}
where we have set the notation $\op{\phi}\coloneqq \op{\phi}_I(z, \tau)$ for the time dependent fields in the interaction picture. The second term in the integrals here is given by the subtraction of vacuum bubbles. Performing the appropriate Wick contractions on this expression gives
\begin{align}
    \langle\op{A}^2&\op{B}^2\rangle_{\Omega}\nonumber
    \\
     &=
    \langle \op{A}^2\rangle_0\langle \op{B}^2\rangle_0 - \frac{i\lambda}{2}\langle\op{B}^2\rangle_0\int d\tau dz \langle\mathcal{T}\op{A}\op{\phi} \rangle_0^2\langle\mathcal{T}\op{\phi}^2\rangle_0 
    \nonumber
    \\
    &~~-
    \frac{i\lambda}{2}\langle\op{A}^2\rangle_0\int d\tau dz \langle\mathcal{T}\op{B}\op{\phi}\rangle_0^2\langle\mathcal{T}\op{\phi}^2\rangle_0 
    \nonumber
    \\
    &
    ~~- i\lambda \int d\tau dz \langle\mathcal{T}\op{A}\op{\phi} \rangle_0^2\langle\mathcal{T}\op{B}\op{\phi} \rangle_0^2.
    \label{firsttermcovarianceinteractions}
\end{align}
Evaluating the second term in Eq.~\eqref{interactingquadraticfieldcovariance} gives the expression 
\begin{align}
    \langle\op{A}^2\rangle_\Omega &\langle\op{B}^2 \rangle_{\Omega}
    \nonumber
    \\
    &= \left(\langle \op{A}^2\rangle_0 - \frac{i\lambda}{2}\int d\tau dz \langle\mathcal{T}\op{A}\op{\phi} \rangle_0^2\langle \mathcal{T}\op{\phi}^2\rangle_0 + \mathcal{O}(\lambda^2)\right)
    \nonumber
    \\
    &~~\times
    \left(\langle \op{B}^2\rangle_0 - \frac{i\lambda}{2}\int d\tau dz \langle\mathcal{T}\op{B}\op{\phi} \rangle_0^2\langle \mathcal{T}\op{\phi}^2\rangle_0 + \mathcal{O}(\lambda^2)\right)
    \nonumber
    \\
    &=
    \langle \op{A}^2\rangle_0\langle \op{B}^2\rangle_0 - \frac{i\lambda}{2}\langle \op{B}^2\rangle_0 \int d\tau dz \langle \mathcal{T}\op{A}\op{\phi}\rangle_0^2\langle \mathcal{T}\op{\phi}^2\rangle_0 
    \nonumber
    \\
    &~~-
    \frac{i\lambda}{2}\langle \op{A^2}\rangle_0\int d\tau dz \langle \mathcal{T}\op{B}\op{\phi}\rangle_0^2\langle\mathcal{T}\op{\phi}^2\rangle_0 +\mathcal{O}(\lambda^2).
\end{align}
Subtracting this from Eq.~\eqref{firsttermcovarianceinteractions} gives the full covariance between $\op{A}^2$ and $\op{B}^2$ in the interacting vacuum as
\begin{align}
    C_{AB} = -i\lambda \int d\tau dz \langle\mathcal{T}\op{A}\op{\phi}\rangle_0^2\langle\mathcal{T}\op{B}\op{\phi}\rangle_0^2 + \mathcal{O}(\lambda^2).
\end{align}
Up to order $\mathcal{O}(\lambda)$, this is simply the contribution from a tree level diagram with two external points $\op{Q}^n_j$ and two external points $\op{Q}^m_{h}$. Given that the operator $\op{\phi} \coloneqq \op{\phi}_I(z, \tau)$ here has full Fourier support over the reals and is not localised to any one wavelet scale, the quantities $\langle\mathcal{T}\op{A}\op{\phi}\rangle_0$ and $\langle\mathcal{T}\op{B}\op{\phi}\rangle_0$ will have overlapping support with $\op{\phi}$ within $\chi_n$ and $\chi_m$, respectively. For the equal time vacuum, we can Wick rotate $\langle\mathcal{T}\op{A}\op{\phi}\rangle_0$ and $\langle\mathcal{T}\op{B}\op{\phi}\rangle_0$ to Euclidean time and represent them as their Fourier Kernels as
\begin{subequations}
\begin{align}
    \langle\mathcal{T}\op{A}\op{\phi}\rangle_0 \rightarrow  \mathcal{K}^n_j &\coloneqq \bra{0}\op{Q}^n_j(0)\op{\phi}(z, s)\ket{0}_E,
    \nonumber
    \\
    &~= \sqrt{h_n}\int_\reals \frac{dk}{2\pi}\chi_n(k)e^{ik(x_{n,j} - z)}\frac{e^{-\omega_k|s|}}{2\omega_k},
    \\
    \langle\mathcal{T}\op{B}\op{\phi}\rangle_0 \rightarrow \mathcal{K}^m_h &\coloneqq \bra{0}\op{Q}^m_h(0)\op{\phi}(z, s)\ket{0}_E
    \nonumber
    \\
    &~= \sqrt{h_m}\int_\reals \frac{dk}{2\pi}\chi_m(k)e^{ik(x_{m,h} - z)}\frac{e^{-\omega_k|s|}}{2\omega_k}
\end{align}
\end{subequations}
where $s = i\tau$ is the usual Wick rotated Euclidean time. Thus, to leading non-vanishing order of $\lambda$, the covariance between $\left(\op{Q}^n_j\right)^2$ and $\left(\op{Q}^m_h\right)^2$ is given by
\begin{align}
    C^{nm}_{jh} &=-\lambda\int ds dz \left[\mathcal{K}^n_j(z, s)\right]^2\left[\mathcal{K}^m_h(z, s)\right]^2
    \nonumber\\
    &\quad+\mathcal{O}(\lambda^2),\qquad n\neq m.
\end{align}
We can evaluate the integrals over $z$ and $s$ to give $C^{nm}$ entirely in terms of the Fourier representations of the field as the integral over $z$ gives a Dirac delta function in external momenta and the integral over $s$ gives the reciprocal of the sum over the $\omega_k$. As such, we have
\begin{align}
    C^{nm}_{jh} &= \frac{-\lambda h_n h_m}{8}\int_{\reals^3}\frac{dk_1dk_2dk_3}{(2\pi)^3}\frac{\chi_n(k_1)\chi_n(k_2)\chi_m(k_3)}{\omega_1\omega_2\omega_3\omega_4}
    \nonumber
    \\
    &~~~\times 
    \frac{\chi_m(-k_1-k_2-k_3) \cos\bigl((k_1+k_2)(x_{nj}-x_{mh})\bigr)}{\omega_1+\omega_2+\omega_3+\omega_4}
    \nonumber
    \\
    &~~~+
    \mathcal{O}(\lambda^2).
    \label{fourierrepquadraticcovariance}
\end{align} 
We can see from the numerator of the integral in Eq.~\eqref{fourierrepquadraticcovariance} that the covariance between the quadratic powers of the fields at two different scales has non-zero contributions when $|-k_1-k_2-k_3|\in (2^m\Omega, 2^{m+1}\Omega)$ with $|k_1|, |k_2| \in (2^n\Omega, 2^{n+1}\Omega)$ and $|k_3|\in (2^m\Omega, 2^{m+1}\Omega)$, respectively. Pairs $(n,n,m,m)$ will always admit momenta that satisfy these conditions no matter the scale separation between $n$ and $m$. This is in line with both the infinite range of the quartic coupling strength for a given pair of wavelet scales as seen in Fig.~\ref{fig:interactioncouplingplotnsep} as well as the scale separability condition in Eq.~\eqref{scaleseparationrangecondition}.
Finally, it is useful to note that the same quantity can be determined in the dyadic rational lattice representation with $C_{\eta\eta'} = C^{nm}_{jh}$.  

\section{Outlook \label{sec:waveletsoutlook}}
\subsection{Results summary and interpretation}
Throughout this work, we have used Shannon wavelets to extend the continuous--discrete equivalence provided by Shannon sampling theory beyond fields possessing a single ultraviolet cut-off. Rather than restricting the full field to be bandlimited, we decompose its spatial momentum support into an infinite collection of non-overlapping frequency bands, with each band associated with a different wavelet scale. Each individual scale is thus bandlimited and possesses its own natural sampling lattice, while the collection of all scales is sufficient to reconstruct the original field without imposing an ultraviolet or infrared cut-off. In this sense, a $d$-dimensional continuous field may be represented by a discrete system carrying both position and scale labels, giving an effective $d+1$-dimensional lattice representation. Importantly, this correspondence remains an exact change of representation so long as the wavelet basis is not truncated; ultraviolet and infrared cut-offs may subsequently be introduced, if desired, simply by restricting the range of wavelet scales.

For square-integrable functions, we showed that this decomposition supplies discrete equivalents to several familiar continuous operations. The $L^2$ inner product becomes a sum over wavelet samples, while spatial derivatives are represented by matrices acting separately within each wavelet scale. The lack of overlap between the Fourier support of Shannon wavelets at different scales is particularly important here. It causes the derivative matrices to be diagonal in the scale label and, when the same construction is applied to the free Klein--Gordon field, causes the Hamiltonian itself to decompose into a sum of independent scale Hamiltonians. Consequently, each wavelet component of the free field evolves independently of the others. This same scale separation is visible in the commutation relations and in the vacuum two-point correlation function, for which correlations between distinct scales vanish exactly.

We further showed that the position and scale labels of this lattice may be combined into a single dyadic-rational co-ordinate $\eta$. In this representation the dyadic valuation, or equivalently the dyadic magnitude $|\eta|_2$, takes over the role of the wavelet scale. The resulting description makes the multiscale structure of the theory more geometrically apparent: field samples at lattice sites of equal dyadic magnitude belong to the same momentum shell, while lattice sites of differing dyadic magnitude represent different resolutions of the field. For the free theory, the second-derivative operator, Hamiltonian evolution and two-point correlation function all respect this same decomposition into constant-$|\eta|_2$ sectors. The dyadic-rational representation therefore does more than simply relabel the usual wavelet indices; it makes the relation between scale, lattice density and coupling structure visible within a single co-ordinate system.

This complete scale separation is no longer present once interactions are introduced. For the quartically self-interacting scalar field, products of four wavelets need not be orthogonal between scales, and the interaction Hamiltonian therefore generates couplings between field samples belonging to different wavelet bands. We characterised these interactions using the quartic coupling tensor $\mathcal{E}$. In the usual position--scale representation this is an eight-index object, while on the dyadic-rational lattice it becomes a function of four dyadic co-ordinates. In either representation, conservation of momentum together with the compact Fourier support of the Shannon wavelets imposes a strong selection rule on which combinations of scales can couple. This produces an interacting lattice that is nonlocal in ordinary position but nevertheless highly structured: the coupling strengths decay with real separation, while their admissibility and magnitude are also strongly controlled by separation in wavelet scale, or equivalently by the dyadic magnitudes of the lattice sites.

The form of these couplings is particularly suggestive when contrasted with the free theory. At quadratic order the scale direction is completely disconnected, whereas a polynomial interaction joins otherwise independent scale sectors together. However, these connections are not arbitrary. Momentum conservation forces the interaction network in the scale direction to be sparse, with large separations in scale excluded for some configurations while paired high- and low-frequency contributions can remain admissible over arbitrarily large scale separations. The scale direction introduced by the wavelet transform therefore possesses a nontrivial interaction structure that is inherited directly from the kinematics of the original continuous field, rather than being placed into the lattice theory by hand.

A perhaps less obvious result of this work is that the existence of such inter-scale interactions does not imply a non-zero two-point function between scales. Translation invariance ensures that the exact interacting two-point function remains diagonal in spatial momentum. Since two distinct Shannon scales have disjoint momentum support, their projected two-point function must therefore vanish. This statement does not rely on a perturbative expansion and remains true in the fully interacting vacuum. Thus, despite the presence of interactions, only the diagonal (same scale) components of the Wavelet decomposed two-point function network are non-zero. 

The inter-scale structure of the interacting theory instead becomes visible through higher-order correlation functions. We found that an $N$-point function projected onto Shannon wavelets can only be non-zero when momenta can be chosen from the Fourier support of the $N$ external wavelets such that their total momentum vanishes. This produces a general scale-admissibility condition whose $N=2$ case gives exact scale diagonality of the two-point function and whose $N=4$ case reproduces the selection rule derived independently from the quartic coupling tensor. The same condition places a logarithmic bound on how far an isolated highest-frequency external field can lie from the next-highest scale for a fixed number of external fields. Thus, although the interacting Hamiltonian couples different resolutions of the field, the order of a correlation function itself constrains the range in scale over which that interaction can be observed. This provides a direct relation between momentum conservation, correlation order and multiscale connectivity.

\subsection{Extensions, literature context, and future work}
Wavelets have previously been used in QFT for several related purposes. Wavelet bases provide a natural multiresolution encoding for quantum simulation, with observables and excitations associated with different physical resolutions \cite{Brennen2015}; they have also been used to study the renormalisation of correlations and entanglement across scales \cite{GeorgeBrennen2022}, and there is a close relation between wavelet transforms and entanglement-renormalisation tensor networks \cite{EvenblyWhite2016}. Continuous wavelet approaches have additionally been applied directly to interacting Green functions and the regularisation of field theories \cite{Altaisky2013}. The approach taken here is complementary to these works. In particular, the use of Shannon wavelets gives exactly separated momentum bands, allowing several of the scale-selection rules derived here to follow directly from momentum conservation rather than approximately from localisation or coarse graining. It also extends the continuous--discrete correspondence considered for globally bandlimited fields \cite{UpdatedLewis2023} to a field with unbounded momentum support by resolving that support scale by scale. The discreteness introduced here should therefore be viewed primarily as a choice of basis rather than as a fundamental ultraviolet regulator.

An immediate avenue for future work is to investigate how this formulation interacts with the renormalisation group. The Shannon decomposition already partitions the field into non-overlapping momentum shells, which is very close in spirit to the starting point of Wilsonian renormalisation \cite{Wilson1974}. The difference is that here all shells are retained simultaneously as explicit degrees of freedom. It would therefore be interesting to ask whether integrating out subsets of the wavelet lattice can reproduce a conventional renormalisation-group flow \cite{Fries2019,MichlinPolyzouBulut2017}, and whether the interaction tensor $\mathcal{E}$ provides a useful discrete description of how effective couplings flow between scales. 

That being said, the scale-selection rules derived here may permit many couplings to be discarded identically before any numerical approximation is introduced. We expect such sparsity to make the Shannon representation useful for both classical \cite{HallidaySuranyi1995} and quantum simulation of interacting field theories, extending the multiscale simulation programme of Ref.~\cite{Brennen2015}. 
The SLAC derivative matrix coupling acts to preserve the momentum space structure of the derivative and eliminate the problem of Fermion doubling. Its use is prevelent throughout works dealing with quantum field theories on lattices \cite{UpdatedDrell1974,UpdatedDrell21976} and it has roots in Shannon theory \cite{UpdatedLewis2023}. This work extends on this advantage for simulating fields on lattices by showing that a separation of momentum into different Shannon wavelet scales produces this SLAC discrete derivative coupling for bandpass-type momentum support. The extension of the lattice to have a scale direction preserves the continuous structure of the field by and can act to encode ultraviolet and infrared cutoffs where deired and the wavelet derivative matrices preserve scale diagonality. For interacting fields, a treatment in terms of simulating field observables via correlation functions complements the SLAC derivative structure in terms of scale diagonality. For a field on a lattice indexed by position and time, two-point functions in the interacting vacuum remain scale local while general N-point functions are sparse along scale. 

That being said, there are several technical questions that would need to be addressed before such an approach could be used as a general numerical method. The exact correspondence developed in this work requires infinitely many lattice sites and infinitely many scales. Any practical calculation must therefore truncate both real-space extent and scale range \cite{MichlinPolyzouBulut2017,Basak2026}. Although such a truncation gives ultraviolet and infrared cut-offs in a particularly transparent way, it would be useful to quantify the resulting error in observables and correlation functions and to determine how counterterms should depend on the chosen scale truncation \cite{EliasMiroRychkovVitale2017}. In particular, the perterbative corrections to the two-point function introduced in Sec.~\ref{sec:interactingcorrelationfunctionswavelets} indicate that changing basis does not affect the interacting time evolution unitary and thus does not change the usual ultraviolet structure of the interacting continuum theory. An explicit calculation of these correction terms, including tadpole, sunset, and other corrections via contributing Feynman diagrams would thus be useful for investigating the use of these wavelets for renormalisation and general quantum field theoretic calculations. Additioanlly, a systematic treatment of renormalisation on the truncated Shannon lattice, and a comparison with more conventional momentum-shell and lattice regularisations \cite{Wilson1974,ZarateThiemann2025}, would be a natural extension of the present work.

It would also be worthwhile to investigate interactions beyond the $1+1$-dimensional scalar $\phi^4$ model considered here. The underlying wavelet decomposition itself extends straightforwardly to additional spatial dimensions, but the geometry of the momentum bands and the corresponding scale-selection rules become richer. Fermionic and gauge fields are particularly interesting examples \cite{HaegemanSwingleWalterCotlerEvenblyScholz2018,AltaiskyKogutSusskind2020}. For gauge fields, one would additionally need to determine how gauge constraints and redundancies \cite{KogutSusskind1975} should be represented on a lattice whose nonlocal couplings arise from exact sampling rather than from a conventional finite-difference derivative. More generally, the present construction decomposes spatial momentum and is therefore not manifestly Lorentz covariant. Understanding whether an analogous scale-resolved representation can be formulated covariantly, perhaps through the spectrum of spacetime differential operators \cite{Pye2023} rather than a purely spatial Fourier transform, remains an important open problem.

Finally, the interpretation of wavelet scale as an additional direction naturally raises questions concerning holography. Connections between wavelets, multiscale tensor networks and emergent bulk geometries have been explored in exact holographic mappings and related constructions \cite{LeeQi2016,ChingHuaLee2017,SinghBrennen2016}. More recently, local correlations of wavelet-transformed quantum fields have been used to construct emergent bulk metrics for free scalar and fermionic theories, with AdS-like geometries arising in appropriate limits \cite{VedlBrennen2026}. The discrete construction presented here suggests a complementary question: what geometry, if any, is encoded by the interaction graph of the Shannon lattice once the field is no longer quadratic? For the free theory, the scale layers are disconnected, while self-interactions generate highly constrained links between them. It would be interesting to determine whether these interaction-induced links can be given a geometric interpretation, or whether higher-order correlation functions \cite{FernandezMelgarejoMolinaVilaplanaTorrenteLujan2019} provide a more natural measure of distance in the scale direction than the two-point function, which we have shown remains exactly scale-diagonal. We stress, however, that the appearance of an additional scale co-ordinate by itself is not sufficient to identify the resulting system with a gravitational bulk theory; establishing such a connection would require substantially more structure than has been considered here.

The central result of this work is therefore not simply that a quantum field can be written in a wavelet basis but that doing so exposes a hierarchy in how different resolutions of that field communicate. Free evolution, derivatives and two-point correlations are separated exactly by scale. Self-interactions connect those scales, but only according to kinematic rules inherited from the momentum support of the original continuum theory. Higher-order correlation functions then provide the observables through which this cross-scale structure becomes visible. Shannon wavelets consequently provide an exact setting in which continuous field theory, discrete sampling and multiscale dynamics can be studied within the same framework. We expect that extending this framework to renormalisation, higher-dimensional and gauge theories, numerical simulation, and interaction-dependent emergent geometries will provide several useful directions for future work.

\begin{acknowledgments}
We thank Lucky K. Antonopoulos for their contributory discussions on this work.
We used OpenAI's ChatGPT (6--6.2 Sol and 6--6.2 Astra at various reasoning levels) to assist with manuscript editing, result validation, and literature contextualisation. We reviewed the AI's feedback and used it to inform our research and writing where appropriate. We take full responsibility for the integrity of the research and for the manuscript's content and accuracy.
We thank OpenAI for providing complimentary access to ChatGPT through its ChatGPT for Academic Researchers program.
This work was supported by the Australian Research Council~(ARC) through the Centre of Excellence for Quantum Computation and Communication Technology (CQC$^2$T; Project No.~CE170100012), Centre of Excellence for Quantum Computer Performance and Integration~(QPI; Project No.~CE260100009), and Discovery Projects (Project Nos.~DP200102152 and DP260104888). N.C.M.\ is supported by an ARC Future Fellowship (Project No.~FT230100571).
AK acknowledges support through the Dieter Schwarz Foundation, the Natural Sciences and Engineering Research Council of Canada (NSERC), the National Research Council of Canada (NRC) and the ARC.
\end{acknowledgments}

\bibliography{refs}
\end{document}